%% file: main.tex
\documentclass[lettersize,journal]{IEEEtran}

\usepackage{amsmath,amsfonts}
\usepackage{algorithmic}
\usepackage{algorithm}
\usepackage{array}
\usepackage[caption=false,font=normalsize,labelfont=sf,textfont=sf]{subfig}
\usepackage{caption}
\usepackage{textcomp}
\usepackage{stfloats}
\usepackage{url}
\usepackage{verbatim}
\usepackage{graphicx}
\usepackage{cite}
\usepackage[breaklinks=true]{hyperref}

\pdfoutput=1

\usepackage{lipsum}
\usepackage{siunitx}
\usepackage{cleveref}
\usepackage{booktabs}
\usepackage{xcolor}
\usepackage{transparent}
\usepackage{xspace} % to handle space after the command
\usepackage{xfrac} % provide nice inline \frac

\usepackage{multirow}   % for fancy tables

\definecolor{mygray}{rgb}{0.35,0.35,0.35}% dark grey

\usepackage{microtype}
\input{notation}
\input{notation_common}

\crefname{figure}{Fig.}{Figs.}
\Crefname{figure}{Figure}{Figures}
\crefname{equation}{Eq.}{Eqs.}
\Crefname{equation}{Equation}{Equations}
\crefname{table}{Table}{Tables}
\Crefname{table}{Table}{Tables}

\begin{document}

\title{Gaussian Process Regression of Steering Vectors\\
With Physics-Aware Deep Composite Kernels\\
for Augmented Listening}

\author{
Diego Di Carlo,~\IEEEmembership{Member,~IEEE,}
Shoichi Koyama,~\IEEEmembership{Senior Member,~IEEE,}
Aditya Arie Nugraha,~\IEEEmembership{Member,~IEEE,} \\
Mathieu Fontaine,~\IEEEmembership{Member,~IEEE,}
Yoshiaki Bando,~\IEEEmembership{Member,~IEEE,}
Kazuyoshi Yoshii,~\IEEEmembership{Senior Member,~IEEE.}

\thanks{
Manuscript received xxx yyy, 2025; revised xxx yyy, 2026; accepted xxx yyy, 2026.
Date of publication xxx yyy, 2026; date of current version xxx yyy, 2026.
This study was supported by 
JST FOREST No. JPMJFR2270,
JSPS KAKENHI Nos. JP23K16912, JP23K16913, JP24H00742, and 24H00748.
ANR Project SAROUMANE (ANR-22-CE23-0011),
and Hi! Paris Project MASTER-AI.
The associate editor coordinating the review of this manuscript and approving it for publication was Prof. XXXX. \textit{(Corresponding author: Diego Di Carlo).}       
}%
\thanks{
Diego Di Carlo, Aditya Arie Nugraha, Yoshiaki Bando are with the Center for Advanced Intelligence Project (AIP), RIKEN, Tokyo 103-0027, Japan (e-mail:
diego.dicarlo@riken.jp).
% ; adityaarie.nugraha@riken.jp)

Shoichi Koyama is with National Institute of Informatics, Tokyo, Japan.
% (e-mail: koyama.shoichi@ieee.org).

Mathieu Fontaine is with LTCI, Télécom Paris, Institut Polytechnique de Paris, France.
% (e-mail: mathieu.fontaine@telecom-paris.fr).

Kazuyoshi Yoshii is with the Graduate School of Engineering, Kyoto University, Kyoto 615-8510, Japan, 
and also with the AIP, RIKEN, Japan.
% (e-mail: yoshii.kazuyoshi.3r@kyoto-u.ac.jp).

The experimental code will be shared upon acceptance.
\\\textit{This work has been submitted to the IEEE for possible publication. Copyright may be transferred without notice, after which this version may no longer be accessible.}
}% <-this % stops a space
% \thanks{Digital Object Identifier xxx}%
% \thanks{Source code available \href{https://github.com/chutlhu/gp-steerer}{https://github.com/chutlhu/gp-steerer}}
}

\markboth{Journal of \LaTeX\ Class Files,~Vol.~XX, No.~XX, January~2025}%
{Shell \MakeLowercase{\textit{et al.}}: A Sample Article Using IEEEtran.cls for IEEE Journals}


\maketitle

\begin{abstract}
This paper investigates
 continuous representations 
 of steering vectors over frequency and microphone/source positions for augmented listening 
 (e.g., spatial filtering and binaural rendering), 
 enabling user-parameterized control of the reproduced sound field.
Steering vectors have typically been used 
 for representing the spatial response of a microphone array as a function of the look-up direction.
The basic algebraic representation of these quantities
 assuming an idealized environment
 cannot deal with the scattering effect of the sound field.
One may thus collect 
 a discrete set of real steering vectors 
 measured in dedicated facilities 
 and super-resolve (i.e., upsample) them.
Recently, physics-aware deep learning methods 
 have been effectively used for this purpose.
Such deterministic super-resolution, however,
 suffers from the overfitting problem
 due to the non-uniform uncertainty 
 over the measurement space.
To solve this problem,
 we integrate an expressive representation 
 based on the neural field (NF)
 into the principled probabilistic framework 
 based on the Gaussian process (GP).
Specifically, 
 we propose a physics-aware composite kernel
 that  models the directional incoming waves and the subsequent scattering effect. 
Our comprehensive comparative experiment 
 showed the effectiveness of the proposed method
 under data insufficiency conditions.
In downstream tasks 
 such as speech enhancement and binaural rendering
 using the simulated data of the SPEAR challenge,
 the oracle performances were attained 
 with less than ten times fewer measurements.
\end{abstract}


\section{Introduction}

\begin{figure}
    \centering
    \vspace{-3mm}
    \includegraphics[trim={0 0 0 0},clip,width=.99\linewidth]{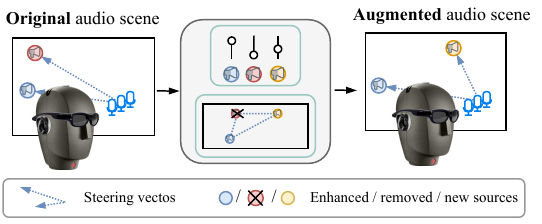}
    \caption{\small A typical \textit{remixing} workflow for augmented listening. The acoustic scene is first analyzed (e.g., DOA estimation and separation) and then spatially remixed before rendering. While semantic content can be modified, spatial information (e.g., steering vectors shown as arrows) must remain coherent to convey realism.}
    \label{fig:augmented-listening}
    \vspace{-3mm}
\end{figure}

\IEEEPARstart{A}{ugmented} listening (AL)~\cite{corey2019microphone} 
 refers to modifying 
 the user's perceived sounds in real time
 for a better auditory experience.
It aims to equally help 
 both normal-hearing and hearing-impaired people
 to hear better what they attend to
 in real noisy echoic situations.
As depicted in \cref{fig:augmented-listening},
 for example,
 personalized sound zones can be produced
 by manipulating the sound field 
 according to the individual preference~\cite{betlehem2015personal}.

Behind such real-time applications, 
 both audio analysis (e.g., localization and separation) 
 and synthesis (e.g., reproduction) 
 have actively been studied
 at the intersection of acoustics, signal processing, 
 and machine learning (ML)~\cite{cobos2022overview}.
Acoustics offers a solid mathematical representation of sound as a \textit{field}, i.e., a continuous function over space and time, and a central goal is to reconstruct this quantity at unobserved locations from finite measurements~\cite{koyama2025physics}.
In signal processing, 
 sound is typically assumed to be generated 
 from a specific source and modified by a filter
 representing the sound propagation (e.g., in room acoustics and reverberation) and spatial information (e.g., source location and array geometry).

Since estimating sources and propagation filters from microphone observations is ill-posed, recent work combines classical array processing with data-driven (deep) learning for joint analysis and synthesis~\cite{cobos2022overview,lee2022differentiable}. However, many state-of-the-art (SOTA) methods still rely on idealized sound-propagation models  (e.g., free-field or simplified room models) to generate training data or design inductive bias.

In the speech enhancement 
 for augmented reality (SPEAR) challenge~\cite{tourbabin2023spear}
 using noisy speech recordings obtained with 
 a head-worn microphone array,
 a deep learning-based method ranked first 
 in both the objective and subjective criteria, 
 while a classic baseline 
 the isotropic minimum variance distortionless response (MVDR) beamformer~\cite{hafezi2023subspace}, 
 came second in subjective assessments. 
Both methods rely heavily on pre-measured propagation filters,
 also known as steering vectors (SV)~\cite{van2002optimum}, 
 highlighting their central role in the AL pipeline.

\begin{figure}
    \centering
    \includegraphics[width=.95\linewidth]{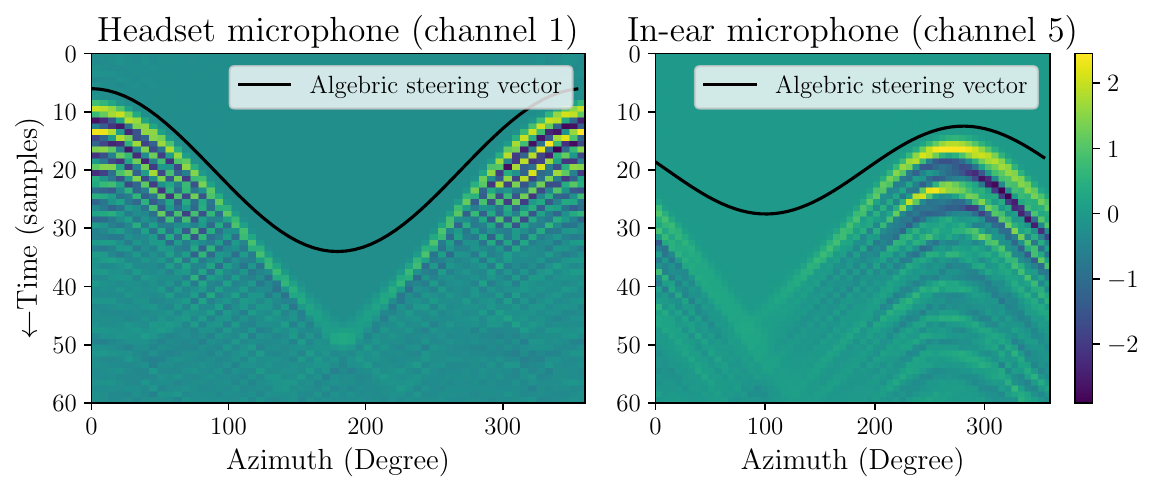}
    \caption{\small Comparison between amplitude of measured and algebraic steering vectors on the azimuthal plane in the time domain, available in the SPEAR Challenge data. }
    \label{fig:svect}
    \vspace{-1em}
\end{figure}

SVs encode the array response
as a function of the direction of the incoming sound~\cite{van2002optimum}.
Their estimation
 has been a key technique
 in spatial audio processing such as
 speech enhancement~\cite{gannot2017consolidated},
 sound source localization~\cite{chardon2022theoretical}, 
 and sound scene synthesis~\cite{zhang2017surround}. 
Specifically, an SV comprises the complex, frequency-dependent transfer functions from a far-field direction to each array microphone. Under this definition, SVs may include both the listener’s head-related transfer functions (HRTFs) as well as multichannel room impulse responses~\cite{gannot2017consolidated}. Their estimation can be viewed as a structured instance of sound-field estimation in which the field is parameterized by direction. The objective of this paper is to learn a continuous (HRTF-like) SV model over direction (and frequency) from sparse measurements using head-worn microphones where dense acquisition is time-consuming, enabling beamforming and rendering for augmented listening.

While continuous HRTF representations have been studied extensively (e.g., spherical-harmonic/subspace-based interpolation),
The continuous nature of SVs naturally calls for
 the use of a \textit{neural field} (NF),
 a deterministic nonlinear function over a continuous domain,
 parametrized by a deep neural network (DNN)~\cite{xie2022neural}. 
In general, 
 a NF is trained from a limited number of SVs
 measured at sampling points (on a grid)
 over space and frequency,
 achieving resolution-free SV upsampling 
 (interpolation).
This approach, however, 
 can neither deal with uncertainty 
 originating from the noisy, sparse SV measurements
 over the continuous domain
 nor guarantee the fidelity at measurement points.
Physics can only be weakly incorporated
 into supervised learning, 
 either through additional losses for regularization
 (physics-informed approach)
 or
 parametrization of the function
 (physics-constrained approach)~\cite{koyama2025physics}.

To overcome these limitations,
 in this paper
 we propose a statistically principled method for SV upsampling
 based on the Gaussian process regression (GPR)
 with NF-based deep kernel learning (DKL).
To compute the covariance between any pair of SVs
 over space and frequency, 
 we define a physics-aware composite kernel
 as the product of multiple NF-based kernels
 corresponding to
 the directional propagation
 and the subsequent scattering.
This kernel is used for formulating a \textit{prior} GP 
 in the noiseless SV space
 strictly obeying the physical laws.
Given noisy SV measurements 
 at sparse sampling points,
 we optimize the kernel parameters,
 i.e., NFs and noise variance,
 such that the likelihood is maximized.
We then compute a \textit{predictive} GP
 that gives the distributional estimates of noiseless SVs
 at arbitrary points including the measurement points.

At the heart of this work
 is a marriage of physics (deduction)
 and statistics (induction)
 in the DKL-GPR framework
 for resolution-free SV upsampling.
A key advantage of our method
 is to deal with the uncertainty
 about the SV function over space and frequency,
 i.e., consider all possible SV functions,
 in both the DKL and GPR,
 achieving robust and interpretable prediction
 of noiseless SVs from noisy sparse measurements.
Another significant contribution of this work is
 demonstrating our algorithm's practical utility
 within an AL pipeline.
Specifically, 
 we show that upsampling the SVs increases 
 the spatial selectivity of an MVDR beamformer, 
 thereby yielding improved speech enhancement performance.

\section{Related Work}\label{sec:sota}

% The SV
%  represents the propagation of sound waves 
%  over a microphone array 
%  as a function of the lookup direction~\cite{van2002optimum}. 
% It is also known as the \textit{array manifold} vector 
%  in antenna array processing~\cite{schmidt1992multilinear} 
%  and can be viewed as a special case 
%  of the \textit{plenacoustic function} 
%  in acoustics~\cite{ajdler2003plenacoustic}.
 
% In free-field anechoic environments, 
%  SVs are represented as a closed-form expression 
%  based on the array geometry~\cite{van2002optimum}.
% This approach offers a simple, fast, differentiable formulation 
%  of SVs with a geometric interpretation.
% The acoustics reflection and sound scattering, however,
%  are typically regarded as noise.
% In non-free-field anechoic environments, in contrast, 
%  SVs should account for additional factors, e.g., 
%  scattering objects and microphone directivity patterns. 
% The HRTF~\cite{xie2013head} is a kind of SV 
%  that captures effects caused by human anatomical features 
%  regarding the pinnae, head, and torso. 
% Such vectors are typically estimated 
%  through in-lab measurements 
%  or computationally heavy simulators~\cite{bruschi2024review}.

% The steering vector (SV)
%  is a function of direction and frequency
%  representing the sound propagation 
%  over a microphone array~\cite{van2002optimum}. 
Steering vector (SV) maps 
 frequency and look direction
 to the complex propagation gains received 
 by a microphone array~\cite{van2002optimum}.
Under free-field conditions, 
 it admits a closed-form geometric expression,
 whereas in realistic environments, 
 room propagation, scattering and microphone directivity 
 must be incorporated~\cite{valimaki2012fifty}.
\cref{fig:svect} depicts the difference between
 algebraic and measured SVs 
 for several directions on the azimuthal plane
 in the presence of a human head 
 as scattering object.

SVs can be numerically simulated 
 based on acoustics 
 using general-purpose acoustic simulators starting from the geometry and acoustics properties of the scattering objects~\cite{brinkmann2023recent}.
Nevertheless, 
 real-world complexity limits their accuracy, 
 motivating data-driven estimation method~\cite{bahrman2024speech}
 and blind separation methods~\cite{gannot2017consolidated,bando2021neural}, 
 albeit still challenged by robustness issues.
 % ~\cite{ratnarajah2023towards}.

% Beyond anechoic conditions,
%  one may numerically compute SVs 
%  using an acoustic simulator~\cite{valimaki2012fifty}.
% The applicability of this approach, however, 
%  is limited by the computational complexity 
%  and the insufficient knowledge about the acoustic environment. 
% Algebraic simulators can be implemented 
%  in a differentiable manner in the deep learning architecture
%  for analysis-by-synthesis estimation~\cite{kelley2024rir} 
%  and virtually supervised learning~\cite{lee2023yet}.
% However, such approaches
%  may still struggle to replicate 
%  real-world complex acoustic profiles~\cite{cobos2022overview}.
% Instead, SVs are often estimated within the frameworks
%  of RIR estimation
%  ~\cite{bahrman2024speech} 
%  and blind source separation~\cite{bando2021neural}.
% Despite progress in this area, 
%  robust estimation remains a significant challenge ~\cite{jalmby2023low,ratnarajah2023towards}.

Alternatively,
 one may directly measure SVs 
 under target conditions.
 % ~\cite{zhang2017surround}. 
However, for ad-hoc devices and personalized configurations, such measurements are often available only at a sparse set of spatial locations due to acquisition time and cost~\cite{rafaely2004analysis}.
% Then, measured vectors are often stored in a lookup table, 
%  making them susceptible to dataset-specific conventions
%  and preprocessing requirements~\cite{zhang2023hrtf}.
Spatial upsampling has thus been studied
 for improving the spatial resolution of SVs,
 especially for
 field reconstruction (SFR)~\cite{koyama2025physics}
 and virtual auditory display with HRTFs~\cite{porschmann2020comparison}.
 
% In AL applications, 
%  anechoic SVs encode the scattering sound field
%  caused by a sound wave impinging a set of sensors, 
%  which are typically embedded in a head-worn headset~\cite{cobos2022overview}.
% The upsampling of such quantity 
%  in the spatial domain is related to \textit{HRTF upsampling}~\cite{bruschi2024review} 
%  and \textit{sound field reconstruction} (SRF)~\cite{koyama2025physics}.
% Besides, modeling the acoustic field 
%  for further multichannel audio analysis 
% is studied in blind source separation 
%  and speech enhancement frameworks~\cite{gannot2017consolidated}.
% \Cref{tab:sota} provides a 
%  schematic summary of the literature 
%  covered in this section.

\subsection{Sound Field Reconstruction}

Sound-field reconstruction (SFR) aims to infer the acoustic field, 
 a continuous function over space and time, at locations where no measurements are available, using observations acquired by a set of microphones.
The target field has been expressed
 with expansions of plane wave~\cite{
 caviedes2021gaussian,ribeiro2024sound,
 bertin2015compressive,koyama2019sparse},
 spherical harmonics (SH)~\cite{pezzoli2022sparsity}, 
 acoustics modes~\cite{das2021room}, 
 or equivalent sources~\cite{antonello2017room}.
These functions are used 
 for regularized linear regression~\cite{pezzoli2022sparsity,koyama2019sparse}
 or kernel ridge regression~\cite{ribeiro2024sound}. 
Besides their theoretical guarantees, 
 these methods have several limitations.
First, some expansions, e.g., SH on the unit sphere, and Bessel functions on a bounded domain,
 rely on orthogonal expansion,
 whose truncation order
 relates to the number of available observations.
Second, these models typically focus on
 reconstructing relatively small zones
 featuring several microphones, 
 which in our survey of representative works (\cite{caviedes2021gaussian,ribeiro2024sound,bertin2015compressive,koyama2019sparse,pezzoli2022sparsity,das2021room,antonello2017room}), are rarely composed of fewer than 5 microphones, with the best performance typically achieved when using more than 16.
 
Deep learning (DL) has recently been a key technique for SFR.
For example, UNet-like architectures 
 were used to estimate the low frequencies 
 of the sound field (up to \SI{300}{\hertz})
 %e.g., in 
 \cite{lluis2020sound,kristoffersen2021deep}.
 % in a pure data-driven fashion. 
Generative models
 such as generative adversarial 
 networks (GANs)~\cite{karakonstantis2023generative}, 
 deep image priors~\cite{pezzoli2022deep}, 
 diffusion models~\cite{miotello2024reconstruction},
 and neural processes~\cite{liang2024sound},
 have been proven to be effective for SFR.
Specifically, the GAN-based method~\cite{karakonstantis2023generative} reports benefits in reconstructing higher-frequency components (up to 4~\si{kHz}).
The pure data-driven nature of these methods, however,
 can suffer from the discrepancy between training and test acoustic conditions.
While acoustic simulation can support data augmentation~\cite{bryan2020impulse},
 their intrinsic accuracy-speed trade-off limits the realism and diversity of generated conditions, leaving unseen real-world acoustics uncovered.

To address this issue, 
 physics-driven ML methods have been proposed.
In general, physics is incorporated as an \textit{inductive bias} 
 for ensuring output structure, 
 overcoming data scarcity, 
 and reducing the opaqueness of deep models.
Physics-driven ML can be further 
 categorized into physics-\textit{informed}~\cite{pezzoli2023implicit,ma2024sound,karakonstantis2024room,ribeiro2024sound} 
 and physics-\textit{constrained} approaches~\cite{karakonstantis2023generative,caviedes2023spatio,ito2022head,feng2024room,bi2024point}.
The former, known as physics-informed neural networks (PINNs), 
 directly represent a target field parametrized by a DNN
 and optimize it
 with a multi-objective loss
 including a regularization term
 evaluating the residual for a PDE.
Such DNNs are hard to train in practice
 due to multiple loss terms~\cite{rohrhofer2023apparent}. 
Besides, 
 the deviation of physics only minimized, 
 meaning that physical properties are not guaranteed at output.
Instead, 
 physics-constrained approaches estimate 
 expansion coefficients for known basis functions 
 with DNNs~\cite{ito2022head,karakonstantis2023generative,bi2024point} 
 or Bayesian models~\cite{feng2024room,caviedes2023spatio},
 demonstrating superiority over traditional methods 
 for interior SFR at low frequencies with several microphones.

This study extends the physics-driven ML approach
 for reconstructing an exterior acoustic field 
 surrounding a human head 
 using only six microphones, 
 up to speech-relevant frequencies (up to $\SI{8}{\kilo\hertz}$). 
Note that most existing models are deterministic 
 and lack uncertainty quantification. 
This has recently been addressed 
 using the GPR~\cite{rasmussen2006gaussian}
 for sound field reconstruction~\cite{caviedes2021gaussian}.
We use the GPR 
 in combination of NF-based DKL for SV upsampling.

\begin{table}[t]
    \centering
        \caption{\small Schematic organization of the selected literature related to acoustic steering vector upsampling.}
    \label{tab:sota}
    \resizebox{\linewidth}{!}{%
    \input{tables/sota_table_small}    }
\end{table}

\subsection{HRTF Upsampling}

HRTF upsampling is a special case of SFR
 featuring the human head and torso as scattering objects. 
Application to binaural rendering 
 motivates two main assumptions.
First, the setup corresponds to
 far-field anechoic binaural in-ear microphones;
second, 
 the focus is usually on interpolating 
 real log-magnitude coefficients of the filters,
 rather than complex-valued spectra~\cite{masuyama2024niirf}.
In contrast to a recent survey of HRTF interpolation~\cite{bruschi2024review} 
 we here describe a complementary taxonomy 
 focusing on data-driven or knowledge-driven methods.
 % with the latter being subdivided 
 % into physics-based and parametric models.
% Authors of~\cite{bruschi2024review} recently reviewed
%  the SOTA techniques for HRTF measurement interpolation.
% Here, 
%  we propose a complementary taxonomy focused on the prior knowledge:
%  we divide the methods into data-driven and knowledge-driven approaches, which are further subdivided into physics-driven and approximated parametric methods.

Classic data-driven methods 
 interpolate local measurements using  
 % bilinear~\cite{begault20003D,freeland2002efficient}, trilinear~\cite{gamper2013head},
 % barycentric~\cite{cuevas20193D}, 
 extension of linear interpolation
 methods (e.g., barycentric or natural neighbor)
 \cite{porschmann2020comparison}.
Subspace methods
 % and wavelets~\cite{torres2009hrtf} 
 have been proposed 
 to reduce the complexity, enabling fast global interpolation
 \cite{xie2012recovery}. 
These methods perform well with dense measurements
 (10–15$^\circ$ spacing),
 but become unreliable with sparser sampling 
 (e.g., 30–40$^\circ$ spacing)~\cite{hogg2024hrtf}.
 % / fewer than 50 points

Recently, data-driven models with DL 
 have been investigated 
 for spatial upsampling~\cite{jiang2023modeling,gebru2021implicit,hogg2024hrtf}
 and extended to other modalities, 
 e.g., anthropometric features and images~\cite{chen2019autoencoding}.
Notably, NF-based methods
 enable HRTF upsampling and representation
 as a function of sound source directions
 \cite {xie2022neural}.
Such an approach has also been used for 
 auralization of audio signals~\cite{gebru2021implicit}
 and unifying HRTFs measured on different spatial grids~\cite{zhang2023hrtf}.

Knowledge-driven approaches address HRTF upsampling
 as a regression problem 
 whose solutions are constrained or regularized by 
 physics-inspired models. 
Classic approaches rely 
 on the geometrical approximation 
 to compute interpolation coefficients~\cite{pulkki1997virtual,luo2013kernel,chen2023head} 
 or spherical interpolation~\cite{zotkin2009regularized}.
Furthermore, DSP-based methods demonstrate 
 the possibility of smoothly interpolating the parameter space 
 of filters modeling the HRTF spectrum~\cite{watanabe2003interpolation,nowak2022spatial} which has been 
 recently extended to use NFs as backend~\cite{masuyama2024niirf}.

Finally, 
 physics-driven methods use the Helmholtz equation 
 (or its parameterized homogeneous solution) 
 to constraint model output~\cite{evans1998analyzing,ito2022head,romigh2015bayesian} 
 or regularize a PINN model~\cite{ma2024sound}.  
% ~\cite{ajdler2008sound,evans1998analyzing,duraiswami2004interpolation,zotkin2009regularized,ahrens2012hrtf,porschmann2019directional,zaunschirm2018binaural,ben2019efficient,arend2021assessing,ito2022head,romigh2015bayesian}
In HRTF upsampling, 
 most of the above methods operate in the SH domain,
 whose expansion coefficients being typically estimated via regularized least-squares~\cite{evans1998analyzing,duraiswami2004interpolation,zotkin2009regularized}. 
However, reconstruction quality is sensitive
 to the measurement grid: 
 while structured grids yield consistent results~\cite{arend2019spatial}, 
 random sparse sampling significantly degrades performance~\cite{ben2019efficient}. 
To address this, 
 methods leveraging neural networks~\cite{ito2022head}, 
 Bayesian inference~\cite{romigh2015bayesian}, 
 time-alignment~\cite{brinkmann2018comparison,ben2019efficient} or directional equalization ~\cite{porschmann2019directional,arend2021assessing}
 have been proposed.
The studies above 
 have addressed upsampling ego-centric 
 acoustic measurements at the two ears. 
In contrast, 
 we investigate the underexplored case of 
 very sparse measurements using  head-mounted microphones. 
While interpolation is often applied independently across frequencies,~\cite{luo2013kernel} proposed a joint spatial-spectral GPR framework. We extend this direction by introducing a physically grounded kernel that incorporates inter-channel dependencies.

\subsection{Array Processing}

% Array processing exploits the geometrical structure 
%  of an array system to address several analysis problems,
%  typically source localization, denoising, and separation. 
In array processing, 
 the \textit{array manifold} maps
 % is the continuous locus 
 % of all the sensors' response vectors 
 %and maps 
 % a geometrical property, e.g., 
 a lookup direction
 to a signal space of SVs~\cite{manikas2004differential}. 
The correct modeling of this quantity 
 enables high-resolution direction of arrival (DOA) estimation
 and source signal enhancement.
The interpolation of array manifold 
 from discrete measurement was addressed 
 in early studies, %~\cite{schmidt1992multilinear}, 
 but due to 
 % the complexity and 
 suboptimal performances, 
 it evolves into a calibration problem~\cite{qiong2003overview}.
If the array geometry is known, 
 SVs can be computed geometrically, 
 % leading to \textit{analytical} manifolds%~\cite{friedlander2017antenna}.
 then, the calibration aims 
 to recover unknown complex gain and phase factors
 to compensate for the mismatch,
 typically using maximum-likelihood-based approaches~\cite{qiong2003overview}.
To the best of our knowledge, 
 no studies address 
 interpolation of array manifold, 
 nor have they used deep learning models
 to model this continuous object.

Finally, 
 some multichannel array processing frameworks 
 for sound source separations
 can be used
 % operating under the narrowband assumption 
 % use Bayesian inference 
 to estimate array responses.
Here, sound propagation impinging on an array is modeled
 through frequency-independent spatial covariance matrices (SCMs), 
 whose principal components may identify with SVs~\cite{markovich2009multichannel},
 which can be parameterized continuously by the source
direction~\cite{bando2021neural,sumura2024joint}.
Alternatively, 
 SVs can represent relative transfer functions (RTFs), 
 describing the ratio between the ATF of two sensors~\cite{gannot2017consolidated}.
Some works in sound source localization and separation
 use manifold learning techniques to define 
 the continuous 
 locus of  RTFs as a function of source direction or location~\cite{duraiswami2005manifolds,grijalva2017interpolation,laufer2020data}.
 
\section{Proposed Method}\label{sec:proposed}

\begin{figure}[t]
\centering
  \includegraphics[trim={0 0 120 0},clip,width=.98\linewidth]{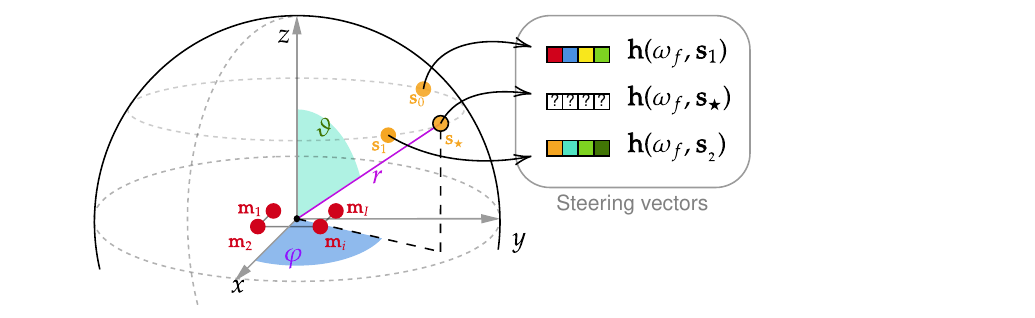}
  \caption{\small Measurement grid and reference system used in this study and illustration of the steering vector interpolation problem.}
  \label{fig:ref-system}
\end{figure}

In the frequency domain, 
 the homogeneous Helmholtz equation describes 
 the evolution of the complex acoustic pressure field $h \in \setC$
 as a function of position $\anyPos \in \setR^3$ 
 and the angular frequency $\omega \in \setR$ as~\cite{ueno2025sound}
\begin{equation}\label{eq:helmholtz}
    \nabla^2 h(\omega,\anyPos) + \frac{\omega^2}{c^2} h(\omega,\anyPos) = 0,
\end{equation}
where 
 $\nabla^2$ is the 3-dimensional Laplacian operator 
 (with respect to space coordinate)
 and $c$ is the speed of sound.

In the presence of a point source, the Helmholtz equation is inhomogeneous at $\srcPos$, while it is homogeneous everywhere else. It's closed form solution is the free-space Green's function (point-source solution).
Thus, assuming a direct-path free space propagation, 
 a single-frequency point source 
 at position $\srcPos \in \mathbb{R}^3$ 
 emits a pressure wave in the form of
\begin{equation}\label{eq:freefield}
    h^d(\omega,\micPos, \srcPos) = \frac{1}{4 \pi r} e^{-\jmath \omega r / c},
\end{equation}
where $r = \| \micPos - \srcPos \|_2$ 
 is the distance between the source 
 and the measurement location, and $\jmath = \sqrt{-1}$. 

The presence of a object (e.g., the user's head) 
 diffracts the sound wave 
 and modifies the pressure field. 
Most of the studies 
 model an observed sound field
 as a sum of an incidental wave 
 and the wave that is scattered around the head~\cite{caviedes2021gaussian}. 
In this paper,
 we express a sound field 
 as a component $h^d$ being modified 
 by the propagation $h^s$ around the head,
 i.e.,
 as filtering~\cite{colton1984inverse}, 
 which in the frequency domain writes as
\begin{equation}\label{eq:prod-model}
    h(\omega, \micPos, \srcPos) = h^d(\omega, \micPos, \srcPos) h^s(\omega, \micPos, \srcPos).
\end{equation}
Equivalently, $h^s$ can be interpreted as time-aligned HRTF, 
while $h^d$ captures the remaining direct-path term\cite{ben2019efficient}.

\subsection{Spatial Sound Field Model}
The acoustic sound field $x(\omega,\micPos,\srcPos)$ 
 measured at the sensor position $\micPos$ 
 produced by a sound source emitting a signal $s(\omega)$ 
 at location $\srcPos$, can be then computed as
\begin{equation}\label{eq:mic-filter-src}
    x(\omega,\micPos,\srcPos) = h(\omega, \micPos,\srcPos) s(\omega).
\end{equation}

Alternatively, the sound field can be represented
 as a linear combination of spatial basis functions~\cite{williams1999fourier}.
In particular, 
 on the spherical surface $\setS^2$,
 centered around the coordinate origin,
 it can be represented by the coefficients $c_{lm}\in\setC$
 of a spherical harmonics (SH) expansion, 
 where $l$ and $m$ are the order and degree of the SH, respectively,
 with basis
 $\Yml(\Doas)\in\setC$
 being a function of the direction $\Doas = (\el, \az) \in \setS^2$,
 where
 $\el$ and $\az$
 are the polar (colatitude) and azimuthal angles, respectively.
The reciprocity principle is applied to the sound pressure of a sound source in the ear of the subject on the surface of a sphere.
Therefore, $ x(\omega, \micPos, \srcPos)$ can be reinterpreted
 as a field radiated from the sensor $\micPos$, 
 which satisfies the exterior Helmholtz equation on the sphere, 
 evaluated at the far field direction $\Omega$.
Since the exterior Helmholtz field restricted to a sphere of fixed radius defines square-integrable boundary data, completeness of the SH $\Yml$ on $\mathbb{S}^2$ yields the following expansion:
\begin{equation}\label{eq:conv-model}
    x(\omega, \micPos, \srcPos) = \sum_{l=0}^{\infty} \sum_{m=-l}^{l} c_{lm}(\omega, \micPos)
        \Yml \left(
        \Doas_{\srcPos,{\arrPos}}
        \right),
\end{equation} 
where 
 $\Doas_{\srcPos,\arrPos}$ 
 is the direction corresponding 
 to the vector $\srcPos - \arrPos$,
 with $\arrPos$ being the array center.
 For a fixed radius at a sufficiently distant location,
 the radial term is absorbed into the coefficients as a constant, which also captures the delay from $\micPos$ 
 and the source’s spectral characteristic.
 
The SH bases of order $l$ and degree $m$ are defined as
\begin{equation}\label{eq:sph-ext}
    \Yml(\Doas) = (-1)^m \sqrt{\frac{(2 l + 1)}{4 \pi} \frac{(l - |m|)!}{(l + |m|)!}} P_l^{|m|}(\cos\el) e^{\jmath m \az}
\end{equation}
where $P_l^{|m|}(\cdot)$ 
 associated Legendre function~\cite{abhayapala2010spherical}.

In practice, 
 the number and location of available measurements 
 limits the maximum SH expansion order $L<\infty$
 which leads to negligible spatial aliasing 
 if $\sfrac{\omega r}{c} \leq L$
 ~\cite{williams1999fourier},
 where $r$ is the radius of the region of interest.
A common rule states 
 that a minimum of $D = (L+1)^2$ directions 
 per frequency 
 is needed to resolve up to order $L$. 

Finally, assuming 
 the source signal 
 to be an ideal impulse
 $s(\omega)=1, \;\forall\, \omega$, 
 the scalar field in \cref{eq:mic-filter-src}
 equals the sum of the directional 
 and scattered scalar fields of~\cref{eq:prod-model}, 
 that is,
 $x(\omega, \micPos,\srcPos) = h^d(\omega, \micPos,\srcPos) h^s(\omega, \micPos,\srcPos)$.

At a sufficiently distant fixed radius, $h^d$ and $h^s$ can be treated as angular functions, allowing them to be expanded using SH as in Eq.~\eqref{eq:conv-model}.
Since $h^d$ and $h^s$ are both 
square-integrable on $\mathbb{S}^2$, their pointwise product is also 
square-integrable on $\mathbb{S}^2$ and, by completeness of $\{Y_l^m\}$, admits the same SH expansion. This is valid boundary data for a unique exterior radiating Helmholtz solution on $\mathbb{S}^2$~\cite{colton1984inverse}. 
Hence, this product represents physically consistent boundary data, corresponding to a physically realizable acoustic field in the exterior domain.

\subsection{Observation Model and Predictions with GP}

Let $y_n := y(\mbz_n)$ denote 
 the $n$-th noisy measurement of the sound field $x$ 
 generated from a single source, such that
\begin{equation}\label{eq:obs-model}
    y_n = x(\mbz_n) + \varepsilon_n, \quad \varepsilon_n \sim \mathcal{N}_{\mathbb{C}}(0, \sigma^2),
\end{equation}
where $\varepsilon_n$ is a zero-mean Gaussian noise 
 with variance $\sigma^2$. 
 $\mbz_n = [\omega_n, \micPos_n, \srcPos_n] \in \setR \times \setR^3 \times \setS^2$ 
 is the collocation point composed of
 space-frequency coordinate,
 as illustrated in \cref{fig:ref-system}.
% The noise term $\varepsilon_n$ models eventual measurement error

Let $\mby = [y(\mbz_1), \ldots, y(\mbz_N)]^\Tr \in \setC^{FJI}$ and $\mbZ = [\mbz_1, \ldots, \mbz_N]^\Tr \in \setR^{N \times 6}$ 
 be the vector of $N = FIJ$ measurements of the sound field
 at $F$ frequencies, $I$ microphone and $J$ source positions, respectively.
Assuming the source to be a perfect impulse $s(\omega) = 1$, 
 the model in \cref{eq:obs-model} writes 
 $y_n = h(\mbz_n) + \varepsilon_n$. 
Thus, given $\mby$ and $\mbZ$, 
 we here focus on the regression problem 
 of estimating the underlying continuous function $h$, 
 that is, the reconstruction of the sound field 
 at another frequency and locations $\mbz_\ast$. 

Following \cite{luo2013gaussian}, 
 we assume the prior distribution 
 over $\mbh = [h(\mbz_1),\ldots,y(\mbz_N)]^\Tr$
 is a complex Gaussian expressed as
\begin{equation}
    \mathrm{p}(\mbh \,|\, \mbZ) \sim \mathcal{N}_\setC (\mbZe, \mbK),
\end{equation}
where $\mbK \in \setC^{N \times N}$ is the Gram matrix 
 whose element $k_{nn'} = k_{\param}(\mbz_n, \mbz_{n'}) \in \mathbb{C}$ 
 is based on the kernel function parameterized by parameters $\param$.
The kernel function $k(\cdot, \cdot)$ models 
 the correlation between points of the underlying function $h$.

The predictive distribution for the test coordinate $\mbz_\star$ is given by
$\mathrm{p}(\mbh_\star \;|\; \mbz_\star, \mby, \mbZ) \sim \mathcal{N}_{\setC}(\mbmu_\star, \mbSi_\star)$,
where the mean $\mbmu_\star$ 
 and the covariance $\mbSi_\star$ are calculated as
\begin{align}
    \mbmu_\star &= \mbk_\star^\Hr (\mbK + \sigma^2 \mbI )^{-1} \mby, \\
    \mbSi_\star &= k_{\param}(\mbz_\star, \mbz_\star) - \mbk_\star^\Hr (\mbK + \sigma_n^2\mbI) \mbk_\star,
    \label{eq:posterior-var}
\end{align}
where $\cdot^\Hr$ denoting Hermitian transposition and $\mbk_\star = [k_{\param}(\mbz_\star, \mbz_1), \ldots, k_{\param}(\mbz_\star, \mbz_n)]$ is the vector of covariances 
 between the test point and the $N$ training points~\cite{rasmussen2006gaussian}.

\subsection{Physics-aware Composite Kernel Design}
In this study,
 we assume that the kernel function 
 % for $\mbh$ 
 can be decomposed as
%  as $\mbK = \mbK^\omega \boxdot (\mbK^d \odot \mbK^s)$ 
%  with $\mbK^\omega \in \setR^{F \times F}$ 
%  and $\mbK^d, \mbK^s \in \setC^{N\times N}$.
% The $\odot$ and $\boxdot$ 
%  are the Hadamard product 
%  and the element-wise product \textit{broadcasted} 
%  over $I$ microphones and $J$ sources, respectively.
% The corresponding kernel reads
\begin{equation}\label{eq:ker}
        k_{\param}(\mbz_{fij},\mbz'_{fij}) =
            k^\omega_{\param}(\omega_f,\omega_{f'})
            k^d_{\param}(\mbz_{fij},\mbz'_{fij})
            k^s_{\param}(\mbz_{fij},\mbz'_{fij}),
\end{equation}
where $\mbz_{fji}' \triangleq [\omega_{f'}, \micPos_{i'}, \srcPos_{j'}]$ is shorthand for readability. 
This model extends the one proposed in~\cite{luo2013gaussian}
 featuring only the $k^\omega$ and $k^s$.

As in~\cite{luo2013gaussian},
 the spectral kernel $k^\omega_{\param}$ is defined as an inverse-quadratic function, 
 yielding an exponentially decaying temporal response 
 with smooth spectral profile: 
\begin{equation}\label{eq:kernel-invq}
    k^\omega_{\param}(\omega_f, \omega_{f'}) 
        = \frac{\alpha}{\ell^2 + (\omega_f - \omega_{f'})^2}, 
\end{equation} 
 where $\alpha$ controls the overall scale 
 and $\ell$ governs the decay rate.
This corresponds to modeling the spectrum
 of a continuous-time auto-regressive AR(1) process~\cite[Appendix B.2.1]{rasmussen2006gaussian}.

The kernel $k^s_{\param}$ models the spatial distribution 
 of the scattering components for fixed microphones. 
 This is derived by the spherical harmonics expansions
 for an exterior sound field
 ~\cite{ito2022head} that is,
\begin{align}\label{eq:ker-src}
    k^s_{\param}(\mbz_{fij},\mbz'_{fij}) 
        &= \Psi(\mbz_{fij})\Psi^\ast(\mbz'_{fij}),\\
    \Psi(\mbz_{fij})
        &= \sum_{l=0}^N \sum_{m=-l}^{l} 
            c_{lm}(\mbz_{fij}) 
            % h_l^1(\omega_f)
            \Yml(\Doas_{\srcPos_j, \anyPos_0}),
            % Y_{lm}(f,(\srcPos_j - \anyPos_0) / R),\\
            \label{eq:spharm-exp}
\end{align}
where $(\cdot)^\ast$ denotes complex conjugate 
 and $\anyPos_0$ is the reference point of the system, i.e., the head center. 
% $\bar{Y}_{lm}(\omega_f,\srcPos_j - \anyPos_0)$ is computed as in~\cref{eq:sph-ext}. 

The kernel $k^d_{\param}$ accounting for the directional components
 is derived from the plane wave kernel~\cite{caviedes2021gaussian}. 
Here, we use SVs as in the definition 
 of rank-1 
 % spatial covariance matrices (SCMs)
 SCMs 
 % in speech enhancement
 \cite{gannot2017consolidated}, 
 that is,
\begin{equation}\label{eq:ker-mic}
    k^d_{\param}(\mbz_{fij},\mbz'_{fij}) = h^d(\omega_f, \micPos_i, \srcPos_j)
    \left(
        h^{d}(\omega_{f'}, \micPos_{i'}, \srcPos_{j'}
    )\right)^\ast,
\end{equation}
where $h^d(\mbz_{fij})$ 
 is the free-field anechoic propagation of~\cref{eq:freefield}.
This kernel models 
 the dominant inter-microphone phase structure 
 predicted by the free-field model for a given frequency 
 and source direction.
Moreover, it accounts for a global delay 
 and the rotation symmetries found in HRTF measurements~\cite{duda1998range}.
This approximation 
 is mainly valid 
 when relative phases are governed 
 by the direct path and is used here 
 to cope with sparse measurements.

\subsection{Parameters Estimation}

\begin{figure}
    \centering
    \includegraphics[trim={0 0 150 0},clip,width=.98\linewidth]{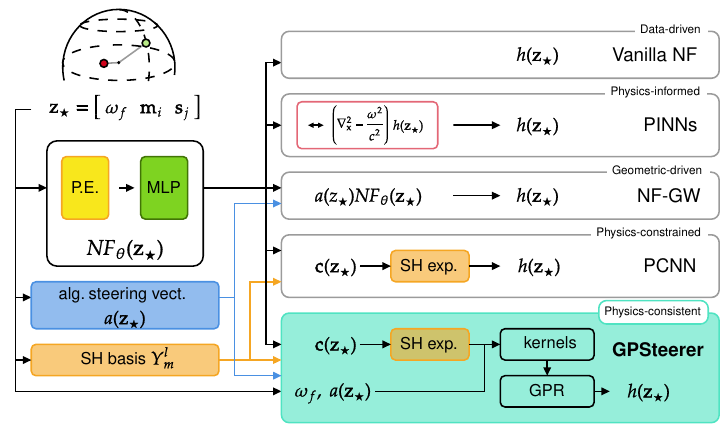}
    \caption{\small Overview of steering vector upsampling models considered in this work. The input $\mbz_\star$ encodes frequency, microphone index, and direction, and $h(\mbz_\star)$ denotes the corresponding complex-valued steering vector. The proposed \texttt{GPSteerer} (in green) uses physics-grounded kernels based on spherical harmonics (SH) expansion, neural field (NF), and algebraic anechoic steering vectors to estimate $h(\mbz_\star)$ using Gaussian-process regression (GPR). Parameter estimation leverages a neural field (NF) model, realized as an MLP conditioned on a positional encoding (P.E.) of $\mbz_\star$.}
    \label{fig:models}
\end{figure}

The parameters to be optimized are
 the ones of the kernel function $k_{\param}(\cdot,\cdot)$ 
 used to compute the prior covariance matrix:
 % of the GP: 
 the decay rate $\ell$, 
 the global scale $\alpha$, 
 the $L$-order complex SH coefficients $\mbc(\mbz_{fij}) = [c_{1}(\mbz_{fij}), \ldots, c_{(L+1)^2}(\mbz_{fij})]^\Tr \in \setC^{(L+1)^2\times 1}$,
 and the noise variance $\sigma^2$.

We propose to use a neural field (NF) 
 with parameters $\param_\text{NF}$
 to estimate the mapping 
 from $\mbz_{ijf}$ to the SH coefficients $\mbc$, that is,
\begin{equation}\label{eq:nf-coeff}
    \mbc(\mbz_{fij}) = \text{NF}(\mbz_{fij}; \param_\text{NF}).
\end{equation}
\cref{fig:models} illustrates the proposed architecture and pipelines,
 featuring sinusoidal positional encoding (P.E.)~\cite{wong2022learning} 
 and an MLP equipped with hyperbolic tangent activation functions,
 together with several variances.

\subsubsection{Loss Function} 

The model parameters are optimized
 by minimizing the following regularized loss function:
\begin{equation}
    \mathcal{L} = - \log p(\mby \,|\, \param_\text{NF}, \ell, \alpha, \sigma^2 ) + \mathcal{L}_\text{reg} ,
\end{equation}
where $p(\mby \,|\,\param_\text{NF}, \ell, \alpha, \sigma^2 ) = \mathcal{N} _{\setC} \left( \mby\,|\,\mbZe, \mbK + \sigma^2 \mbI \right)$ 
 is the likelihood 
 with respect to the model parameters.

The regularization term $\mathcal{L}_{\mathrm{reg}}$ 
 is introduced to enforce smoothness in the spatial domain
 by encouraging sparsity and decay of the spherical harmonics spectrum (SHS)~\cite{pollack1993heat}:
 % corresponding to enforcing smoothness in the spatial domain:
\begin{equation}
 \mathcal{L}_\text{reg} = \lambda_{\ell_1} \sum_{nl} | C_{nl} | + \lambda_\text{exp} \sum_{nl} \text{ReLU}(\mbC_{n(l+1)} - C_{nl}), 
\end{equation}
where the SHS is given by
 $C_{nl} = \sqrt{\frac{1}{2l+1} \sum_m | c_{lm}(\mbz_n) |_2^2 }$.

\subsubsection{Initialization}
To initialize the SH coefficients $c_{lm}$:
 given $D$ measurements, 
 we pre-compute $c_{lm}^{\text{SH}}$ 
 for $l \leq \lfloor \sqrt{D} - 1 \rfloor$ 
 and we sum them to the estimation of the NF, as
\begin{equation}
c_{lm}(\mbz_n) = 
\begin{cases}
    c_{\param,lm}(\mbz_n) + c_{lm}^{\text{SH}}(\mbz_{n}) & \text{if}\:\: l \leq  \lfloor \sqrt{D} - 1 \rfloor,\\
    c_{\param,lm}(\mbz_n) & \text{otherwise},
\end{cases}    
\end{equation}
where $c_{\param,lm}(\mbz_n)$ 
 is the $lm$-th output of the NF in~\cref{eq:nf-coeff}.
Linear interpolation is used 
 to interpolate the pre-computed coefficients $c_{lm}^{\text{SH}}$
 over unseen frequencies.

Prior empirical investigation 
 showed poor reconstruction of the low frequencies
 compared to a SH-based linear regression baseline model
 which achieved almost perfect reconstruction 
 for $f \leq \SI{100}{\hertz}$.
Thus, we use SH-based interpolation 
 to synthetically augment the dataset 
 and pre-train our proposed model for few iterations.

\section{Compared Methods}\label{sec:problem}

\subsection{Linear and Kernel Regression-Based Methods}
A common approach 
 in SFR and HRTF upsampling is 
 to represent a sound field $h(\mbz)$ 
 at space-frequency coordinate $\mbz$ 
 as a linear combination of basis functions, as
\begin{equation}\
    h(\mbz) = \sum_{l=1}^L \gamma _l \sphBasis_l(\mbz) = \mbsphBasis(\mbz)^\Tr \mbga,
\end{equation}
where $\mbga = [\gamma_1, \ldots, \gamma_L]^\Tr \in \setC^L$
 are coefficients and 
 $\mbsphBasis(\mbz) = [\sphBasis_1, \ldots, \sphBasis_L]^\Tr \in \setC^L$
 are the spherical harmonics of \cref{eq:spharm-exp}
 or other basis functions. 
The coefficients $\mbga$ 
 can be computed using linear ridge regression, as
\begin{align}
    \hat{\mbga} &=  \left( \mbPs^\Hr \mbPs + \lambda \mbI \right)^{-1} \mbPs^\Hr \mby,
\end{align}
where 
 $\mbPs = [\sphBasis(\mbx_1), \ldots, \sphBasis(\mbx_N)] \in \setC^{N \times L}$, 
 $\mbI$ is the identity matrix, 
 and $\cdot^\Hr$ denotes Hermitian transposition. 
The above problem is typically ill-posed
 and under-determined which call for
 regularization techniques, 
 such as smoothness 
 or sparsity assumptions~\cite{bertin2015compressive}.
In case of multichannel observations,
 the regression is typically performed 
 for each microphone independently.

Based on the representer theorem~\cite{murphy2012machine}, 
 $h$ can be represented by 
 a weighted sum of kernel functions $k$ as
\begin{equation}
    h(\mbz)  
    = \sum_{n=1}^N \alpha_n k(\mbz, \mbz_n) 
    = \mbk(\mbz)^\Tr \mbal,\label{eq:rkhs}
\end{equation}
where 
 $\mbal = [\alpha_1, \ldots, \alpha_N]^\Tr \in \setC^N$ are the coefficients
 and $\mbk(\mbz) = [k(\mbz, \mbz_1), \ldots, k(\mbz, \mbz_N)]^\Tr \in \setC^{N}$
 is the vector of kernel functions. 
In the kernel ridge regression~\cite{murphy2012machine},
 an estimate of $\mbal$ is computed as
\begin{equation}
    \hat{\mbal} = (\mbK + \lambda \mbI)^{-1} \mby,
\end{equation}
where $\mbK \in \setC^{N \times N}$ is the Gram matrix with elements $k_{nn'} = k(\mbz_n,\mbz_n')$. Then $h$ is interpolated by substituting $\hat{\mbal}$ in \cref{eq:rkhs}.

\subsection{Gaussian Process Regression-Based Method}
GPR is the Bayesian extension to kernel ridge regression.
In~\cite{caviedes2021gaussian},
 GPR for SFR
 involves two spatial kernel functions 
 derived by plane wave decomposition: 
 an anisotropic (i.e., direction-dependent) stationary kernel 
 to model directional components (e.g., direct path and early echoes) 
 and an isotropic stationary kernel for the late reverberation.
In~\cite{luo2013gaussian}, 
 HRTF in both space and frequency
 is assumed to be a zero-mean GP process.
Here, the joint covariance function 
 is defined as the product of an 
 AR(1) process 
 and a stationary covariance 
 based on the  Mat\'ern $3/2$ function
 of the chordal distance expressed as
\begin{align}
    k &= k^\omega(\omega_f, \omega_{f'}) k^s(\Doas_j,\Doas_{j'}), \\
    k^s(\Doas_j,\Doas_{j'}) &= \left(1 + \frac{\sqrt{3} C_{jj'}}{\ell_d} \right) \exp \left(- \frac{\sqrt{3} C_{jj'}}{\ell_d} \right),
    \label{eq:kernel-chmat}
\end{align}
where 
 $k^\omega$ is defined as in~\cref{eq:kernel-invq}, 
 % $\Doas_j = (\az_j,\el_j)$ 
 % is the $j$-th DOA, 
 $C_{jj'}$ is the chordal distance\footnote{$C_{jj'} = 2 \sqrt{\sin^2 \left(\sfrac{\el_j-\el_i}{2}\right) + \sin\el_i \sin\el_j \sin^2\left(\sfrac{\az_i - \az_j}{2}\right)}$ } 
 between DOAs $\Doas_j$ and $\Doas_{j'}$,
 and $\ell_d$ the length-scale.
% is the distance for function values to become uncorrelated in the $d$-th dimension. $\alpha$ and $\lambda$ are the global scale factor and the mean drift rate to $0$ in the OU process, respectively.

\subsection{Neural Field-Based Methods}

A NF
 is a \textit{coordinate}-based neural network
 that maps a point $\mbz \in \setR^d$ 
 to the function value $h \in \setC^l$~\cite{xie2022neural}, for relatively low $d$ and $l$.
The fundamental property of NF is to be grid-free: 
 although the training set is discrete, 
 $\{\mbz_n\}_n \subset \setR^d$, 
 the model can evaluate any point in $\setR^d$. 
 This property enables continuous prediction 
 at inference and for regularization.

Due to spectral bias, 
 standard MLP networks fail 
 to learn a high-frequency function from low dimensional data~\cite{tancik2020fourier}, 
 yielding over-smooth outputs. 
To address this issue,
 two main approaches have been proposed: 
  using random Fourier features~\cite{tancik2020fourier} 
  or sinusoidal activation functions~\cite{sitzmann2020implicit}.
In general, 
 the superiority of the two approaches 
 depends on the specific problem of interest.
 % and on the exploration of the hyper-parameter space. 
The recent work of~\cite{wong2022learning} 
 compared the two approaches 
 against a similar, yet simpler encoding,
 featuring only
% They proposed to transform the 
 % input coordinate into 
 sinusoidal features, as
\begin{equation}
    \mathtt{PE}(\mbz) = [\sin(2\pi\mbW_1 \mbz + \mbb_1)].
\end{equation}
This encoding is then processed by a tanh-MLP backbone, as
\begin{equation}
    \hat{h}(\mbz) = \mathcal{F}_{\param}(\mbz) = \mathtt{MLP}_{\tanh}(\mathtt{PE}(\mbz)).
\end{equation}

NF has been recently applied 
 for continuous representation of 
 sounds~\cite{sitzmann2020implicit},
 acoustics impulse response~\cite{richard2021neural,su2022inras} 
 and HRTF upsampling and personalization~\cite{zhang2023hrtf,masuyama2024niirf,dicarlo2024neural}.
Among these, some studies focus 
 on using geometric properties 
 of far-field propagation 
 to guide the training, 
 also named \textit{geometric wrapping (GW)}~\cite{richard2021neural}, 
 or relying on a parametric model of the HRTF filters~\cite{masuyama2024niirf}.
Besides, 
 each work proposed different MLP configurations 
 (positional encoding, activation functions), 
 % output encoding 
 % (real and imaginary or log-magnitude and phase), 
 and training objectives 
 (magnitude-only loss, combination of magnitude- and phase-based loss terms, etc.) 
 whose performances depend on the specific nature of data.

\subsubsection{Geometric Wrapping} 
GW can be seen as an adaptation of 
 inter-aural phase difference equalization 
 when processing minimum phase HRTF~\cite{kistler1992model}.
The propagation effects 
 at the $i$-th microphone in $\micPos_i$, 
 attending the $j$-th source in $\srcPos_j$
 at frequency $\omega_f$ 
 with respect to the reference $\anyPos$, reads 
\begin{align}\label{eq:svect}
    a(\underbrace{\omega_f,\micPos_i\,|\,\srcPos_j}_{\mbz_n},\anyPos) = \frac{d_{ij}}{d_j} \exp(-\jmath \omega_f (d_{ij} - d_j) / c),
\end{align}
where 
 $d_{ij} = \| \srcPos_j - \micPos_i \|_2$ and $d_j = \| \srcPos_j - \anyPos \|_2$
 are the source distances 
 from the microphone 
 and the references, respectively. 
Finally, the output of the NF 
 featuring GW is given by
\begin{equation}
    \hat{h}_\text{GW}(\mbz_n) = a(\mbz_n,\anyPos)\mathcal{F}_{\param}(\mbz_n).
\end{equation}

\subsubsection{Physics-Informed Neural Networks}
PINNs~\cite{raissi2019physics} are NFs 
 encoding the solution of a PDE 
 which is used as an objective during fitting. 
In case of inverse problems, 
 a data-fit term based on the available measurements is also used.
In the case of frequency-domain SFR,
 the loss function used to optimize
 the  PINN's network parameters
 reads \cite{ma2024sound}
\begin{align}\label{eq:loss-pinn}
    \mathcal{L}_{\param} 
        &= \frac{\lambda_\text{rec}}{N}\sum_{n=1}^N \left( h_n - \hat{h}(\omega_n, \anyPos_n) \right)^2 \nonumber\\
        &\quad+ \frac{\lambda_\text{PDE}}{M} \sum_{n=1}^M \left( \nabla_{\anyPos_n}^2 \hat{h}(\omega_n, \anyPos_n ) + \frac{\omega_n^2}{c^2} \hat{h}(\omega_n, \anyPos_n) \right)^2,
\end{align}
with $\lambda_\text{rec}$ and $\lambda_\text{PDE}$ 
 being hyperparameters regulating the 
 importance of each term.
% Being an extension of NFs models, 
 % a PINN can evaluate any continuous coordinate at test time. 

% When dealing with spherical data (e.g., HRTF), 
%  it is common to adapt polar coordinate systems. 
% Adopting a Cartesian system simplifies the PDE, 
%  automatically handle the wrapping of the spherical data,
%  and avoid coding errors 
%  due to different system representation convention\footnote{This article adopted \href{https://en.wikipedia.org/wiki/ISO/IEC_80000}{ISO 80000-2:2019} 
%  convention as depicted in \cref{fig:ref-system}.}.

The studies in~\cite{karakonstantis2023room,pezzoli2023implicit} 
 proposed a similar approach 
 in the time domain for RIR interpolation, 
 which requires computing the gradient 
 with respect to spatial and temporal coordinates.
The frequency-based modeling is useful 
 to compute GW~\cite{dicarlo2024neural}
 and simplifies the PDE for PINNs~\cite{ma2024sound},
but it needs
 to deal with complex values.
While complex-valued MLP 
 is currently under investigation,
 preliminary studies by the authors showed 
 that returning the real and imaginary part was performing the best.

Training PINNs is known 
 to be challenging in practice~\cite{rohrhofer2023apparent},
 for which several techniques have been proposed, such as,
 meta-learning for multi-objective optimization,
 % ~\cite{wang2022when}, 
 optimizer-switching,
 % ~\cite{dharanalakota2023loss},
 strategic sampling to evaluate the PDE residuals,
 % ~\cite{wu2023comprehensive}, 
 and architecture design choice~\cite{wang2023expert}.
 % ~\cite{wang2022when,luo2022learning}.
% The reader can refer to for a practical introduction'. 
Self-adaptive learning rate annealing, 
 proposed in~\cite{wang2023expert},
 can be used to automatically balance 
 the losses during training.
 % preventing the model from being biased 
 % towards minimizing certain terms during training. 
Specifically, 
 the norm of the gradients of each weighted loss 
 is set to be equal to each other, so that
  $\| \nabla_\theta \mathcal{L}_\text{PDE}\|_2 = \| \nabla_\theta \mathcal{L}_\text{rec}\|_2 = \| \nabla_\theta \mathcal{L}_\text{PDE}\|_2 + \| \nabla_\theta \mathcal{L}_\text{rec}\|_2$, as
\begin{align}
    \lambda_\text{PDE} &= \frac{\| \nabla_\theta \mathcal{L}_\text{PDE}\|_2 + \| \nabla_\theta \mathcal{L}_\text{rec}\|_2}{\| \nabla_\theta \mathcal{L}_\text{PDE}\|_2},\\
    \lambda_\text{rec} &= \frac{\| \nabla_\theta \mathcal{L}_\text{rec}\|_2 + \| \nabla_\theta \mathcal{L}_\text{PDE}\|_2}{\| \nabla_\theta \mathcal{L}_\text{rec}\|_2}.
\end{align}
% \begin{equation}
%     \lambda_\text{PDE} = \frac{\| \nabla_\theta \mathcal{L}_\text{PDE}\|_2 + \| \nabla_\theta \mathcal{L}_\text{rec}\|_2}{\| \nabla_\theta \mathcal{L}_\text{PDE}\|_2},
%     \lambda_\text{rec} = \frac{\| \nabla_\theta \mathcal{L}_\text{rec}\|_2 + \| \nabla_\theta \mathcal{L}_\text{PDE}\|_2}{\| \nabla_\theta \mathcal{L}_\text{rec}\|_2}.
% \end{equation}

At every training iteration, 
 the evaluation of the PDE residual 
 requires sampling the continuous input domain. 
The location and distribution 
 of these residual points 
 impact the training stability 
 and the performance 
 as the model's gradient 
 may vary significantly over the input domain.
A simple, 
 yet an effective approach,
 used in this study,
 consists of using residual points 
 that are uniformly resampled 
 every certain number of iterations~\cite{wu2023comprehensive}.

\section{Evaluation}\label{sec:experiment}

\subsection{Datasets}
We evaluate our method 
 using the SPEAR Challenge dataset~\cite{tourbabin2023spear}, 
 an extension of the EasyCom corpus~\cite{donley2021easycom}, 
 which features real-world egocentric audio-visual recordings
 in dynamic, noisy, and reverberant conditions. 
Data were collected using AR glasses 
 equipped with four microphones, 
 a camera, and binaural in-ear microphones, 
 along with clock-synchronized head pose 
 and close-talking reference signals.

A key limitation of the original EasyCom corpus
 is the absence of ground-truth binaural signals 
 for objective evaluation. 
SPEAR addresses this by providing 
 simulated replicas of the environments, 
 including additional acoustic conditions and conversational setups. 
It also includes anechoic
 acoustic transfer functions (ATFs) 
 for all six microphones (AR glasses and in-ear) 
 over $1020$ directions on a spherical grid
 measured on a head and torso simulator in an anechoic
 chamber. 
We use the manufacturer-supplied array geometry 
 for the glasses and manually calibrated position of the in-ear microphones.
 
\subsection{Experimental settings}

\subsubsection{Task, Metrics, and Data} 

Given a set of SVs 
 observed at $\Nobs \in \{ 8, 16, 32, 64, 128\}$ locations 
  (\textit{train} set), 
  the goal is to estimate the SVs
  for all the directions on the sphere, 
  here represented by the $1020$ available measurements (\textit{test} set). 

As is common in SFR, 
 we consider the following performance metrics. 
The normalized mean squared error (nMSE)
 in decibels per frequency
 captures the reconstruction error
 between the target and estimation 
 and it is computed as~\cite{ueno2025sound}
\begin{equation}\label{eq:nMSE}
    \text{nMSE}(f) = 10 \log_{10} \left( \frac{\sum_{ij} | h_{fij} - \hat{h}_{fij}|^2}{\sum_{ij} |h_{fij}|^2} \right) \quad [\text{dB}].
\end{equation}
To quantify the phase reconstruction 
 in the time domain 
 and the spatial similarity of the filters,
 we use the cosine similarity 
 between estimated and target filters 
 for each direction $j$,
\begin{equation}\label{eq:csim}
    \text{CSIM}(\Omega_j) = \frac{1}{I} \sum_{i} \frac{\sum_t \eta_{tji} \hat{\eta}_{tji}}{\sum_t \eta_{tji}^2 \sum_t \hat{\eta}_{tji}^2} \quad \in[-1,1],
\end{equation}
where 
 $\eta_{tji}$ and $\hat{\eta}_{tij}$ 
 are the time-domain representation of $T$ samples
 of $h_{fij}$ and $\hat{h}_{fij}$, respectively.

\begin{figure}
    \centering
    \subfloat[\label{fig:mollweide}]{{\includegraphics[width=.98\linewidth]{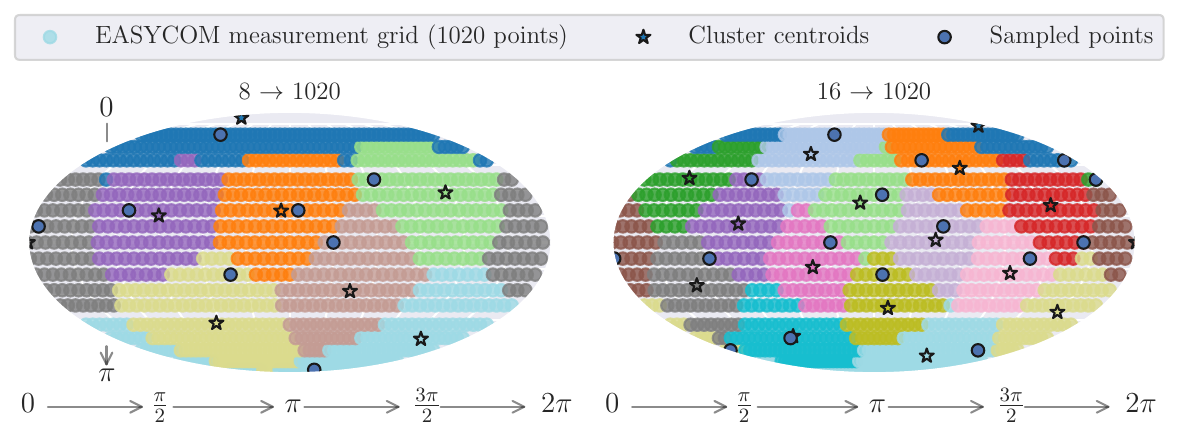}}}%
    \hfill
    \subfloat[\label{fig:valid-split}]{{\includegraphics[trim={0 0 170 15},clip,width=.90\linewidth]{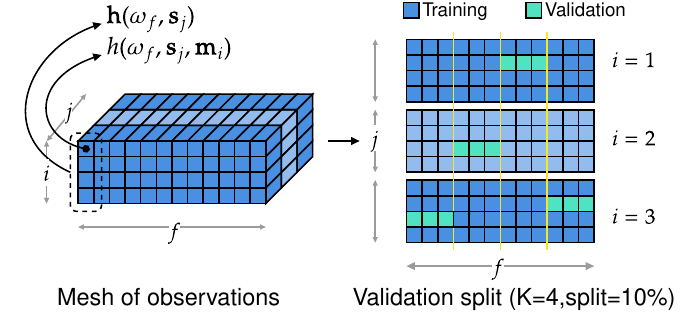}}}%
    \caption{\small \cref{fig:mollweide} shows the Mollweide projection of observed coordinates for two upsampling factors against the observation coordinates in the EasyCom dataset. Stars and colors denote clusters and centroids for a quasi-uniform sampling. \cref{fig:valid-split} illustrates the sampling strategy to select the validation data.}
    \label{fig:enter-label}
    \vspace{-1em}
\end{figure}

Training observations were sampled
 on the unit sphere using 
 the following procedure, 
 depicted in~\cref{fig:mollweide}.
The $1020$ evaluation directions 
 were first clustered into regions 
 using a $\Nobs$-point Fibonacci spherical grid
 as centroids.
Then, 
 one direction was randomly selected from each cluster, 
 except for the frontal cluster, 
 where the fixed point $\Doas = (0,\pi)$ 
 was always included. 
We think this sampling strategy approximates 
 users' measurements
 and mitigates bias in downstream spatial filtering tasks, 
 where users typically face their interlocutors.
Finally, $10\%$ of these measurement
 is used as a held-out validation set 
 for model selection.

% In a scenario where only a few directional measurements are available, $10\%$ of the observation in a held-out fashion for model selection is restrictive, while k-fold cross-validation is demanding for iterative algorithms such as NF. 

For NF-based models, 
 the training data are reshaped 
 to support continuous regression over 
 frequency, microphone, and source position. 
This introduces an imbalance 
 as the spectral axis is denser than the spatial ones.
To mitigate this, 
 we first subsample the frequency axis
 by a factor of two. 
The resulting set is then partitioned 
 into eight equal subsets per direction 
 and channel (see~\cref{fig:valid-split}), 
 from which $10\%$ of the points are randomly selected.
% The number of partitions was empirically optimized.

\subsubsection{Configuration of Compared Methods} 

We compare our models against three baseline interpolation methods: 
 nearest neighbors (\NN) from SciPy library, 
 regularized spherical cubic splines (\SP)
 % \cite{perrin1989spherical} 
 from the MNE library\cite{gramfort2013meg}, 
 and regularized spherical harmonics (\SH). 
 The hyperparameters 
 (\SP: smoothness $10^{-5}$, 50 Legendre terms, stiffness 3; 
 \SH: smoothing $10^{-5}$, order $L = \lfloor \sqrt{\Nobs} - 1 \rfloor$)
 were tuned on the held-out validation set. 

All neural field
 variants (standard \NF, 
 physics-informed \PINN, 
 SH-based physics-constrained \PCNN, 
 and a geometry-warped \NFGW)
 share the same architecture: 
 a 3-layer MLP with 128 hidden units, 
 $\tanh$ activations, 
 and a 128-dimensional sinusoidal positional encoding. 
Input coordinates are non-dimensionalized~\cite{wang2023expert} 
 and scaled by the gain vector 
 $\mbg = [g_f, g_{\srcPos_x}, g_{\srcPos_y}, g_{\srcPos_z}, g_{\micPos_x}, g_{\micPos_y}, g_{\micPos_z}] = [10, 1, 1, 1, 1, 1, 1]$ 
 to balance resolution across different dimensions.

In terms of model capacity, the NF-based variants (\NF, \NFGW, \PINN, \PCNN) have on the order of $10^4$ trainable parameters, whereas the effective parameterization of the \SP{} and \SH{} baselines grows with the number of available observations (approximately $10^3\times N_{\text{obs}}$). 
In contrast, the GP-based baseline \GPChmat involves only 5 kernel parameters. The complete breakdown across models and upsampling factors is reported in the supplementary material.

Training used a batch size of $B = 1024$ 
 and the Adam optimizer with a base learning rate of $10^{-5}$, 
 preceded by a linear warm-up from $10^{-4}$ over 1000 steps, 
 and followed by exponential decay (rate 0.9 every 1000 steps, with a floor at $10^{-5}$). 
Empirical tuning showed that
 gradient clipping at 1
 and disabling weight decay improved performance.
 % as indicated in~\cite{wang2023expert}. 
 All models were implemented in JAX
 % \footnote{\url{https://github.com/jax-ml/jax}}
 % ~\cite{bradbury2018jax} 
and complex-valued SH were computed with 
 a JAX-based differentiable implementation
 % \footnote{\url{https://sphericart.readthedocs.io/}}.
 \cite{bigi2023sphericart}.

\subsubsection{Configuration of Proposed Model}\label{sec:results-par:configuration}
The proposed model, \Proposed, 
 follows the same architectural setup 
 as the NF-based baselines 
 but differs in its objective: 
 instead of directly regressing the steering vector, 
 it predicts a parameterization of the kernel function. 
Consequently, 
 the NF parameters must align
 with the kernel's functional structure.
The best-performing configuration 
 uses a 2-layer MLP, 
 a learning rate ramping from $10^{-4}$ to $10^{-3}$, 
 and input coordinate gains set to 
 $\mbg_f = \mbg_\srcPos = 1$, and $\mbg_\micPos = 100$. 
Prior experiments showed that
 performances are relatively insensitive 
 to the gains on frequency and source position, 
 likely due to the kernel’s modeling. 

Overall, the proposed model 
 has fewer than $9\times 10^4$ trainable parameters, 
 i.e., about $2.5\times$ larger than the NF-only baselines, 
 yet smaller than \SH{} and \SP{} when $N_{\text{obs}}=128$.

% The proposed model, \Proposed, 
%  was configured similarly 
%  to the NF-based models discussed above. 
% While the latter performs a direct regression 
%  in the steering vector space, 
%  the \Proposed predicts a 
%  parameterization of the kernel function.
% Therefore, 
%  the positional encoding must match
%  this space's underlying structure. 
% We found the following parameterization 
%  to perform the best: 
%  learning rate starting 
%  and peak values are set to $10^{-4}$ and $10^{-3}$, respectively; 
%  the number of layers of the MLP is 2; 
%  the gain for each input coordinates is $\mbg_f = \mbg_\srcPos = 1, \mbg_\micPos = 100$. 
% Prior investigations 
%  show that the performances 
%  do not change significantly 
%  by modulating the gain parameter 
%  for the frequencies and source position axes. 
% This is probably due to the kernel formulation, 
%  which allows the control of the spectral bias along these axes. 
%  By contrast, the microphone position coordinates 
%  require higher input gain to match the different number coordinates along this axis.
% To compute the SH-based kernel,
%  we used the differentiable complex-valued implementation available in~\cite{bigi2023sphericart}.
% Ablation study

\begin{figure*}
    \centering
    \includegraphics[width=.98\linewidth]{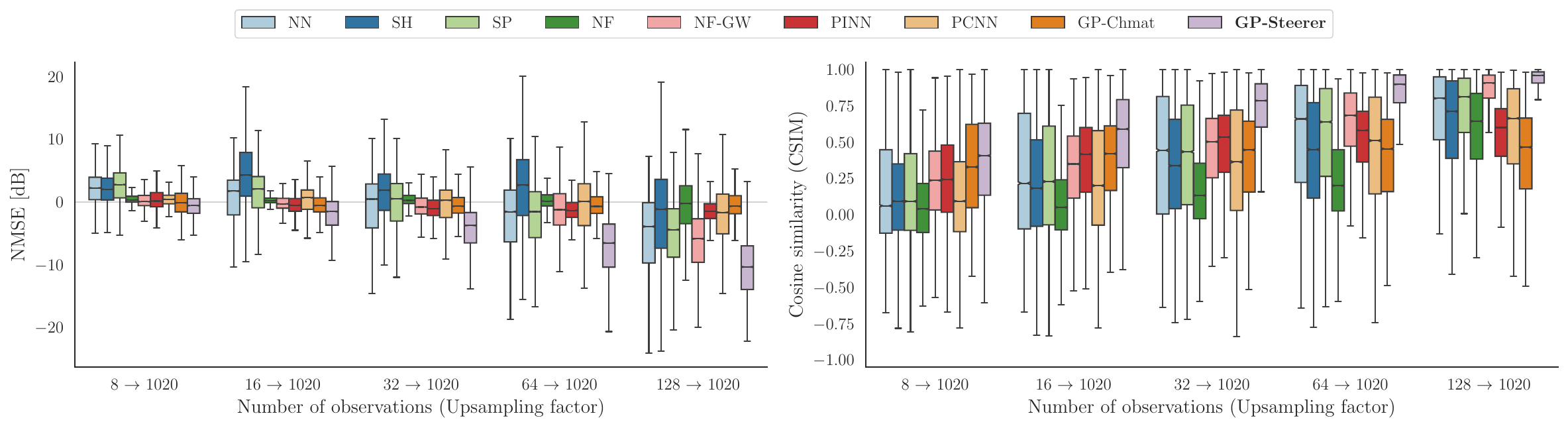}
    \caption{Interpolation results: normalized mean squared error (left) and cosine similarity per number of observed directions.}
    \label{fig:nMSE-csim-avg}
\end{figure*}

\begin{figure}
    \centering
    \includegraphics[width=.95\linewidth]{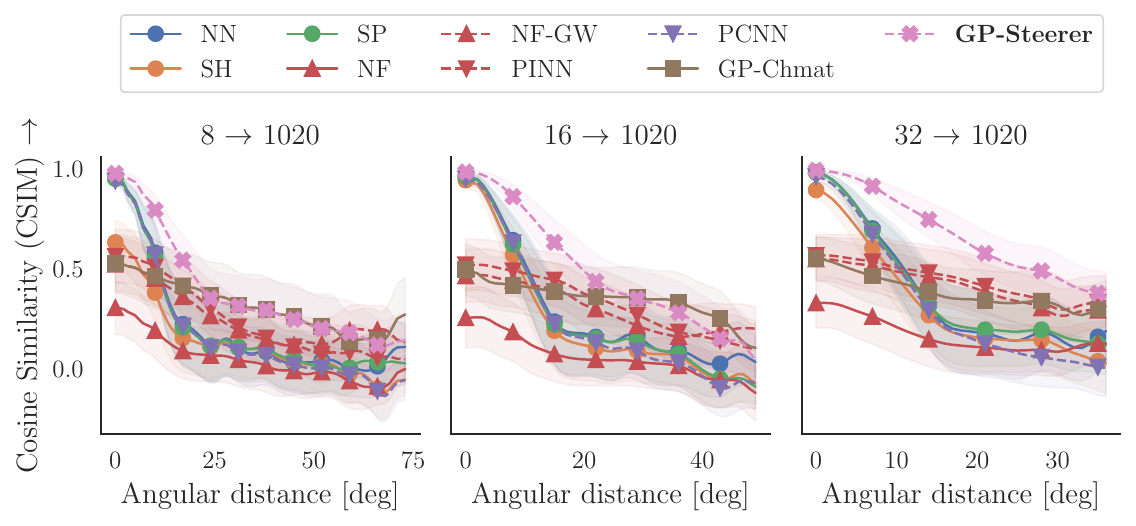}
    \\
    \includegraphics[trim={0 0 0 50},clip,width=.95\linewidth]{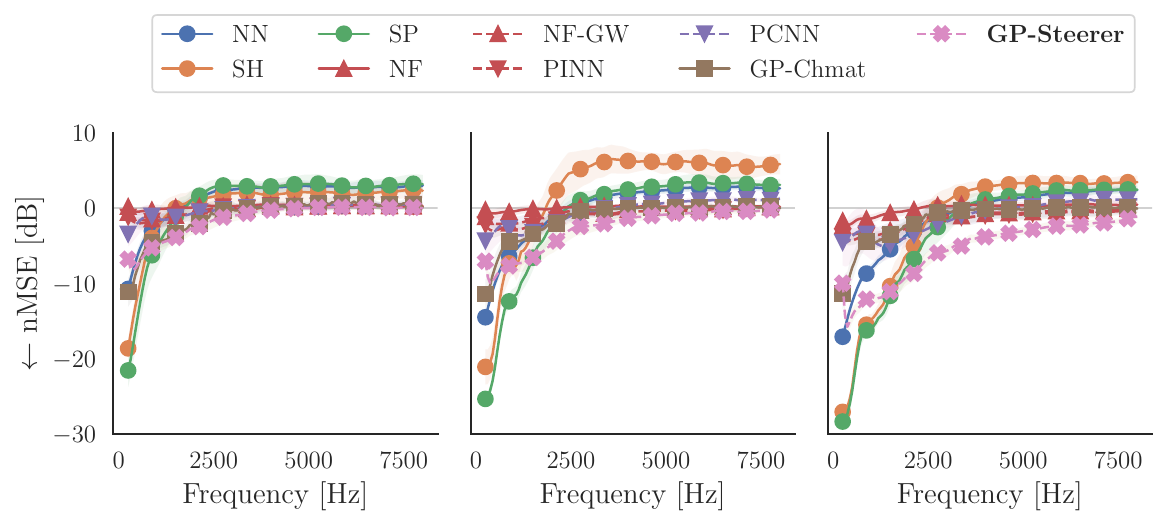}
    \caption{\small Interpolation results: (top) Cosine similarity (CSIM) versus (angular) distance from an observed sample. (bottom) Normalized Mean Squared Error (nMSE) in dB versus frequency. }
    \label{fig:interp-adist-freq}
    \vspace{-1em}
\end{figure}

\subsection{Steering Vector Upsampling}

In this section, 
 we compare the proposed method \Proposed
 for steering vector spatial and spectral upsampling 
 against the three classical baselines described earlier (\NN, \SP, \SH), 
 four NF-based approaches (\NF, \NFGW, \PINN, \PCNN) 
 and a GP regression-based method (\GPChmat). 
For each upsampling factor, 
 results are averaged over
 3 random samplings
 of source positions on the sphere.

\cref{fig:nMSE-csim-avg} reports the averaged 
 interpolation results in terms of nMSE and CSIM.
As expected, 
 the performances of all the methods decrease
 with the sparsity of the observations. 
Nonetheless,    
 it is clear to see the benefit 
 of the proposed method over the compared methods, 
 both in low and higher spatial sampling regimes.
Quantitatively, 
 for the densest setting ($128\rightarrow 1020$), 
 \Proposed achieves a median nMSE of approximately $-13$~dB 
 and a median CSIM of approximately $0.95$, 
 whereas the strongest baselines remain around $-11$~dB and $0.85$. 
In the sparsest setting ($8\rightarrow 1020$), 
 all methods cluster near $0$~dB nMSE with CSIM typically below $0.4$.
 
Among the baseline methods, 
 \SP yields better spatial interpolation results 
 as the performance of \SH
 is affected by the non-regularity of measurements' location. 
Being a purely data-driven approach trained on limited data, the \NF struggles to produce accurate approximations, although its interpolation ability improves with denser observations. 
Incorporating prior knowledge (e.g., the GW) 
 enhances spatial coherence, but yields limited gains in nMSE,
 denoting a priority towards 
 physically plausible solutions
 over pointwise accuracy.

The \PINN exhibits a similar trend: 
 while its PDE-based regularization 
 improves spatial coherence, 
 it performs worse than baselines in terms of nMSE.
This likely stems from the challenge
 of balancing data- and physics-driven losses 
 in multi-objective optimization.
Moreover, since the PDE regularization
 lacks boundary conditions, 
 the solution may drift toward anechoic solutions.
Similarly, \GPChmat benefits from smoothness priors
 and physical constraints at low sample counts, 
 but also saturates, 
 probably due to the
 limited expressiveness
 of the kernel function.

\begin{figure*}[th!]
    \centering
    \includegraphics[trim={0 0 0 40},clip,width=\linewidth]{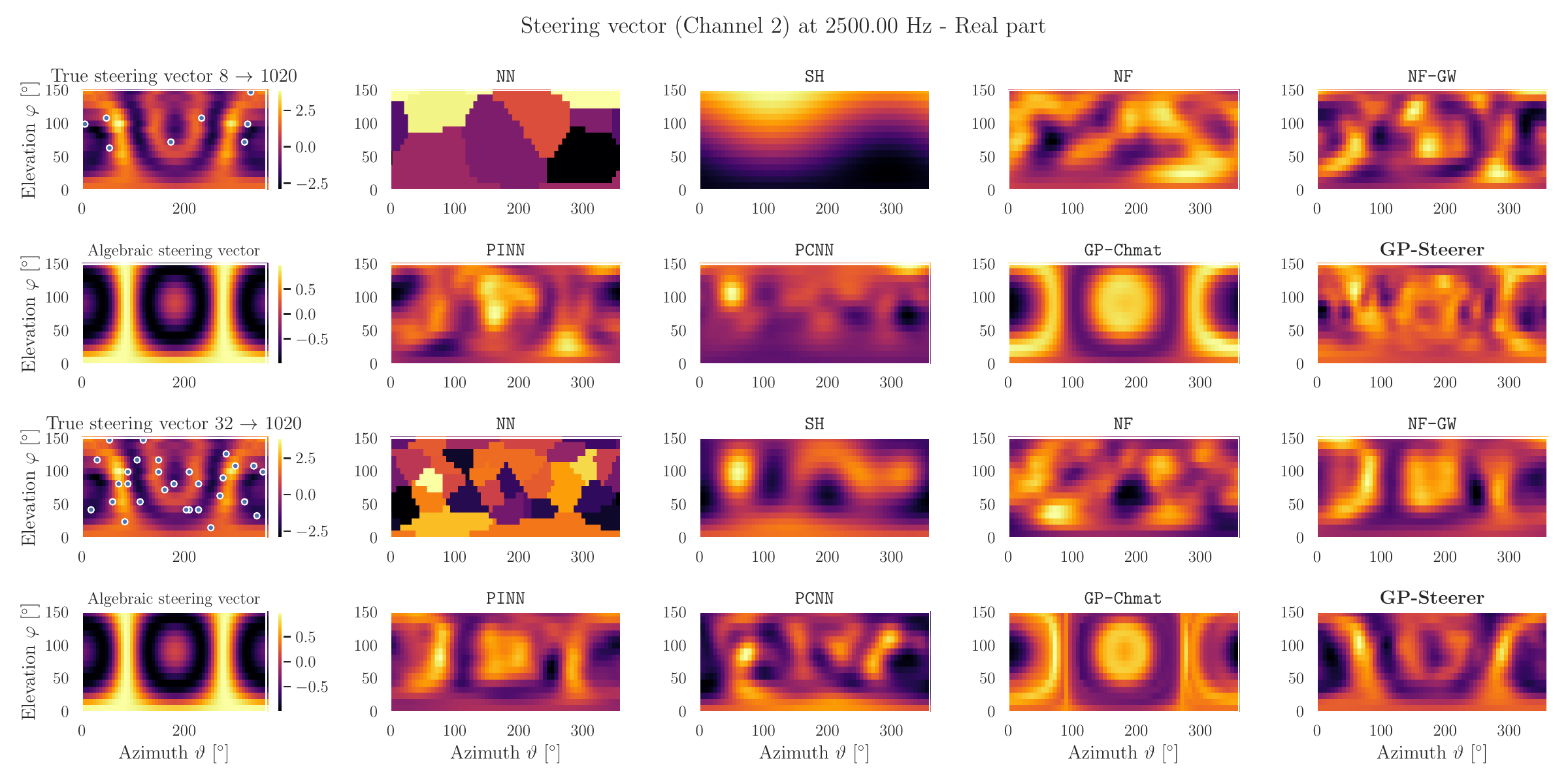}
    \caption{\small Qualitative results of interpolation: real part of the reconstructed steering vector at channel 2 at $\SI{2.5}{\kilo\hertz}$ for 2 sampling factors and different methods. A part for the algebraic steering vectors, all values are normalized by the the maximum range of the ground-truth data. The steering vector are computed with the algebraic model.}
    \label{fig:interpolation-qualitative}
\end{figure*}

\begin{figure*}[t]
    \centering
    \includegraphics[%
    trim={15 15 20 20},clip,%
    width=.98\linewidth]{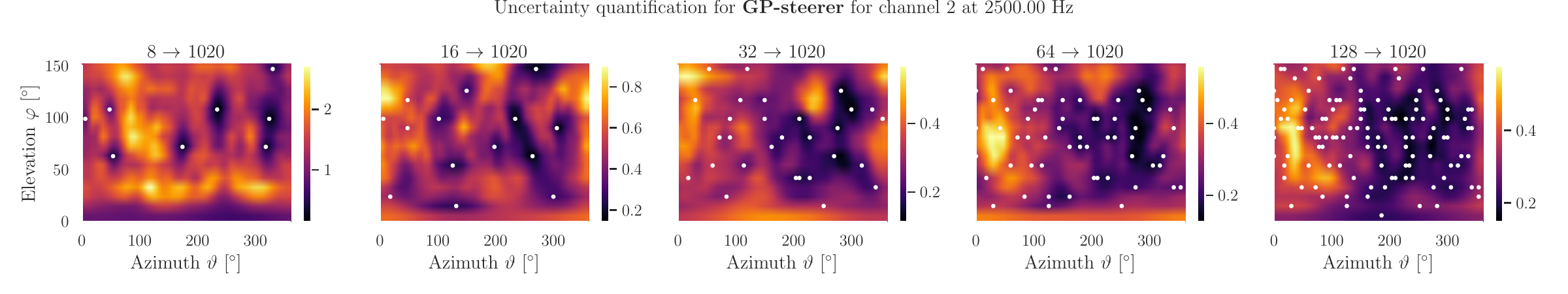}
    \caption{\small Uncertainty quantification as standard deviation of the predicted steering vector at channel 2 at $\SI{2.5}{\kilo\hertz}$ of the proposed model (\Proposed) for different upsampling factors. White dots denote measurement locations.}
    \label{fig:interpolation-post-variance}
\end{figure*}

Additional insights 
 are shown in \cref{fig:interp-adist-freq} (top), 
 where CSIM is plotted against angular distance 
 $\Delta\alpha_{jr} = 2 \arcsin(C_{jr}/2)$, with $C_{jr}$ as in~\cref{eq:kernel-chmat}, between each test direction $j$ and its nearest training direction $r$.
Two regimes emerge: 
 local methods (\NN, \SH, \SP) preserve nearby observations 
 and achieve high CSIM ($\approx 0.8$) 
 for $\Delta\alpha < \SI{10}{\degree}$, 
 making them suitable for downstream tasks 
 like frontal beamforming. 
However, 
 their performance rapidly degrades with distance. 
In contrast, 
 learning-based models with embedded priors 
 (\NFGW, \PINN, \GPChmat) 
 generalize better at larger angular separations,
 denoting some degree of generalization,
 though performing poorly on reconstructing
 known measurements.
The proposed method combines the strengths 
 of both approaches: 
 its GP formulation preserves observed measurements 
 for accurate local interpolation,
 while its physics-informed kernel 
 enables competitive reconstruction at distant, 
 unseen locations.

The large variance 
 observed in the results in \cref{fig:nMSE-csim-avg}
 mainly reflects frequency-dependent performance: 
 low-frequency bins are reconstructed more accurately 
 than high-frequency bins at low values of $N_\text{obs}$, 
 and pooling them yields broader distributions.
To make this explicit, we report
 the nMSE averaged 
 across configurations over positive frequency bins 
 in \cref{fig:interp-adist-freq} (bottom).
As expected, 
 performance degrades with frequency, 
 reflecting the spatial resolution limits
 governed by sampling density\cite{williams1999fourier}. 
The \SH baseline performs well
 at low frequencies—interpolating below \SI{2}{\kilo\hertz} 
 with nMSE better than \SI{-15}{\decibel} 
 using 32 observations, but overfits at higher frequencies, introducing spurious artifacts (positive nMSE in dB). 
\SP yields similarly strong low-frequency performance without such artifacts. 
Other methods, including NF- and GP-based models, typically reach \SI{-10}{\decibel} at best. 
While suboptimal in relative terms, 
 these results align with prior work:\cite{koyama2025physics} reports comparable performance above \SI{1.4}{\kilo\hertz}, 
 and\cite{caviedes2021gaussian} reports values in the range $[-10, 5]~\si{\decibel}$ for sparse-field reconstructions.

The proposed model surpasses 
 all baselines above \SI{2}{\kilo\hertz}, 
 confirming its advantage in high-frequency reconstruction.
Below this threshold, 
 the error remains under \SI{-10}{\decibel},
 highlighting the benefit of the initialization with low-order SH.
Interestingly, 
 all methods perform poorly beyond \SI{2}{\kilo\hertz}. 
Due to the cost of constructing full covariance matrices, 
 GP-based methods (\GPChmat and \Proposed) 
 operate with only $F = 127$ frequencies,
 yet still interpolate effectively across the spectrum, 
 thanks to the smooth spectral kernel.

To illustrate the above findings, 
 \cref{fig:interpolation-qualitative} reports
 the real part of the frequency-domain 
 SVs for one channel at $\SI{2.5}{\kilo\hertz}$
 for two upsampling factors. 
It can be noticed that baseline \NN, \SH, \NF,
 fails to capture the underlying nature of the 
 measurements.
Meanwhile,
 incorporating physical or geometric priors as in
 \NFGW, \PINN and \PCNN
 improves alignment with the algebraic ground truth
 with 32 measurements.
\GPChmat offers a qualitatively better match 
 by leveraging the algebraic steering vectors as kernels, 
 though its chordal-distance kernel 
 fails to capture the complexity of scattering effects.
In contrast, 
 the proposed \Proposed model employing a mixture of SH-based kernels, effectively balancing data- and physics-driven cues. 
This approach produces a good balance
 between a data- and physical-driven solution
 that helps, even in the challenging scenario of only 8 directions.

Finally, 
 the proposed model inherits from the GPR framework
 the ability to perform uncertainty quantification (UQ). 
\cref{fig:interpolation-post-variance} illustrates the
 UQ of the predicted SV at channel 2 and \SI{2.5}{\kilo\hertz}, 
 across different upsampling factors. 
For each direction, SV's UQ can be identified as 
 the standard deviation, which can be derived from~\cref{eq:posterior-var}~\cite{rasmussen2006gaussian}.
As the number of input measurements increases, 
 the model's uncertainty significantly decreases, highlighting its ability to provide more confident predictions with denser sampling. 
High uncertainty is localized in regions distant 
 from the input directions, 
 particularly in sparse configurations, and progressively vanishes with better spatial coverage. 
This demonstrates 
 the model’s capacity to yield principled uncertainty estimates.

\subsection{Downstream Task: Speech Enhancement} 
In this section, 
 we evaluate SV interpolation methods 
 in the context of speech enhancement 
 using the SPEAR challenge datasets. 
The goal is 
 to enhance the binaural image 
 of a target speaker 
 in realistic acoustic scenes,
 characterized by head motion and overlapping speech. 
In this study, we assume oracle DOAs are given
 and focus on the effect of SV interpolation on enhancement quality.

\subsubsection{Metrics} 
We selected a subset of metrics 
 from the SPEAR challenge 
 to evaluate enhancement performance. 
Energetic metrics include 
 signal-to-distortion ratio (SDR), 
 signal-to-artifact ratio (SAR), 
 and image-to-spatial distortion ratio (ISR), 
 and frequency-wise segmental SNR (fwSegSNR), all expressed in dB.
Speech quality is assessed 
 via PESQ (range $[0,5]$), 
 modified binaural STOI (MBSTOI, $[0,1]$).
As mentioned in ~\cite{tourbabin2023spear},
 MBSTOI and fwSegSNR 
 demonstrated
 strong correlation with subjective ratings. 
All scores are reported 
 as improvements over the unprocessed baseline (\texttt{passthrough})
 and computed using the official evaluation script~\cite{tourbabin2023spear}.

\subsubsection{Data}
Each audio signal in the datasets
 is a 1-minute long excerpt at \SI{48}{\kilo\hertz}, 
 which we downsampled to \SI{16}{\kilo\hertz}, 
 to match the maximum frequency of the proposed SV interpolation models.
Following the evaluation framework of the challenge, 
 the mixtures are transformed in the STFT domain 
 using Hamming windows of $\SI{16}{\milli\second}$ and $50\%$ overlap.
Our research utilizes 
 the simulated datasets from the SPEAR Challenge (datasets D2, D3, D4), 
 which offer reference binaural recordings 
 for objective evaluation. 
These datasets differ in acoustics conditions 
 and amount of voice overlap.

\subsubsection{Beamformer Design}
Let 
 $\mbx(f,t) = [x_1(f,t), \ldots, x_I(f,t)]^\Tr \in \setC^{I\times 1}$ 
 denotes the vector of observed signals $x_i(f,t)$ at time frame $t$, frequency index $f$, microphone index $i$. The beamformer output is
\begin{equation}
    r(f,t) = \mbw^\Hr(f) \mbx(f,t),
\end{equation}
where $\mbw \in \setC^{I\times 1}$ is the beamformer weights. For notation simplicity, $\mbh(f,\Doas) = [h(\omega_f,\mbm_i, \Doas)]_i \in \setC^{I\times 1}$ will denote the narrow-band steering vector pointing at DOA $\Doas$, and $(f,t)$ will be omitted unless specified.

The weights of the MVDR beamformer can be derived as~\cite{gannot2017consolidated},
\begin{equation}
    \mbw = \inv{(\mbd^\Hr \inv{\mbR} \mbd)} \inv{\mbR} \mbd,
\end{equation}
where $\mbd = \mbh(\Doas_s) \in \setC^{I\times1}$ is the steering vector for the target DOA $\Doas_s$, $\mbR \in \setC^{I\times I}$ is the noise covariance whose invertibility and numerical stability are ensured via a frequency-independent diagonal loading, i.e., $\mbR+\rho\mbI$, where $\rho$ is selected according to the condition-number~\cite{hafezi2023subspace}.

The Isotropic-MVDR (Iso-MVDR), also known as a super-directive beamformer, assumes a stationary spherically isotropic noise spatial covariance matrix (SCM) ~\cite{gannot2017consolidated} and is the baseline method in the SPEAR challenge. The associated SCM writes \cite{hafezi2023subspace}
\begin{equation}
    \mbR = \sum_{j \in \mathcal{J}} w_j \mbh(\Doas_j) \mbh^\Hr(\Doas_j),
\end{equation}
where $w_j$ is the quadrature weight for each DOA given by
\begin{equation}
    w_j = \frac{2 \sin\el}{N_\az N_\el} \sum_{m=0}^{N_\el/2 - 1} \frac{\sin((2m + 1) \el_j)}{2 m + 1}
\end{equation} 
where $\el_j$ is the inclination of the $j$-th DOA, and the number of azimuths and inclinations are $N_\az$ and $N_\el$, respectively. 

The SPEAR challenge baseline corresponds to 
 \NNOracle using the SVs at all the $1020$
  direction $\mathcal{J}$ to compute the $\mbR$ 
 and to interpolate over the SVs.
For any other upsampling method, 
 $\mbR$ is constructed 
 using SVs resolved at the oracle full-grid locations $\mathcal{J}$.
Since $\mbR$ is independent of source position and time, it can be precomputed.

\begin{figure*}[t]
    \centering
    \raisebox{-0.5\height}{
    \includegraphics[width=0.9\linewidth]{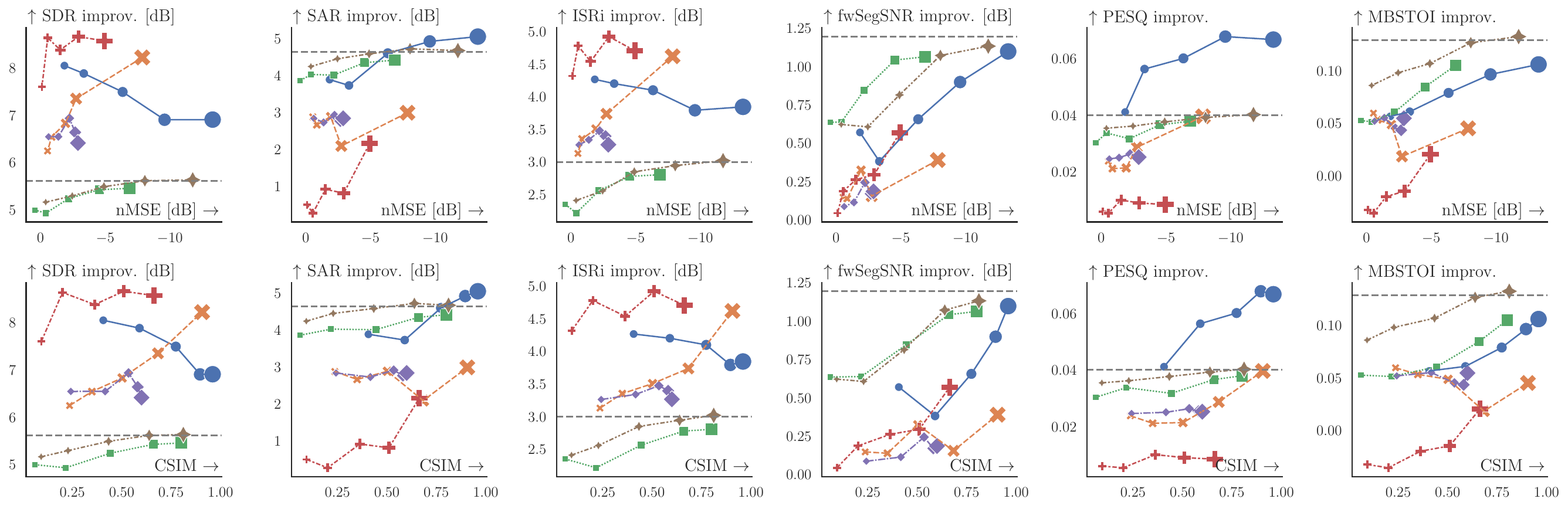}
    }%
    \raisebox{-0.5\height}{
    \includegraphics[trim={0, 0 0 0},clip,width=0.09\linewidth]{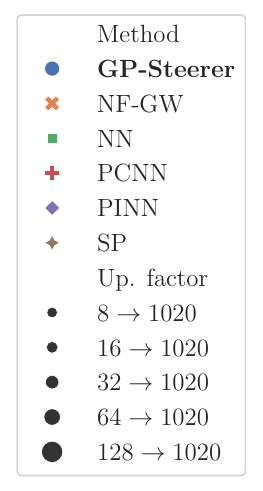}
    }
    \caption{\small Relationship between interpolation quality (nMSE, CSIM) and speech enhancement performance (SDR, SAR, ISR, fwSegSNR, PESQ, MBSTOI) across selected methods. Enhancement metrics are reported as improvements over the unprocessed baseline. Marker size reflects the upsampling factor. Horizontal line denotes oracle methods \NNOracle.}
    \label{fig:metrics-correlation}
    \vspace{-1em}
\end{figure*}

\begin{figure}[t]
    \centering
    \includegraphics[trim={0 30 0 0},clip,width=.90\linewidth]{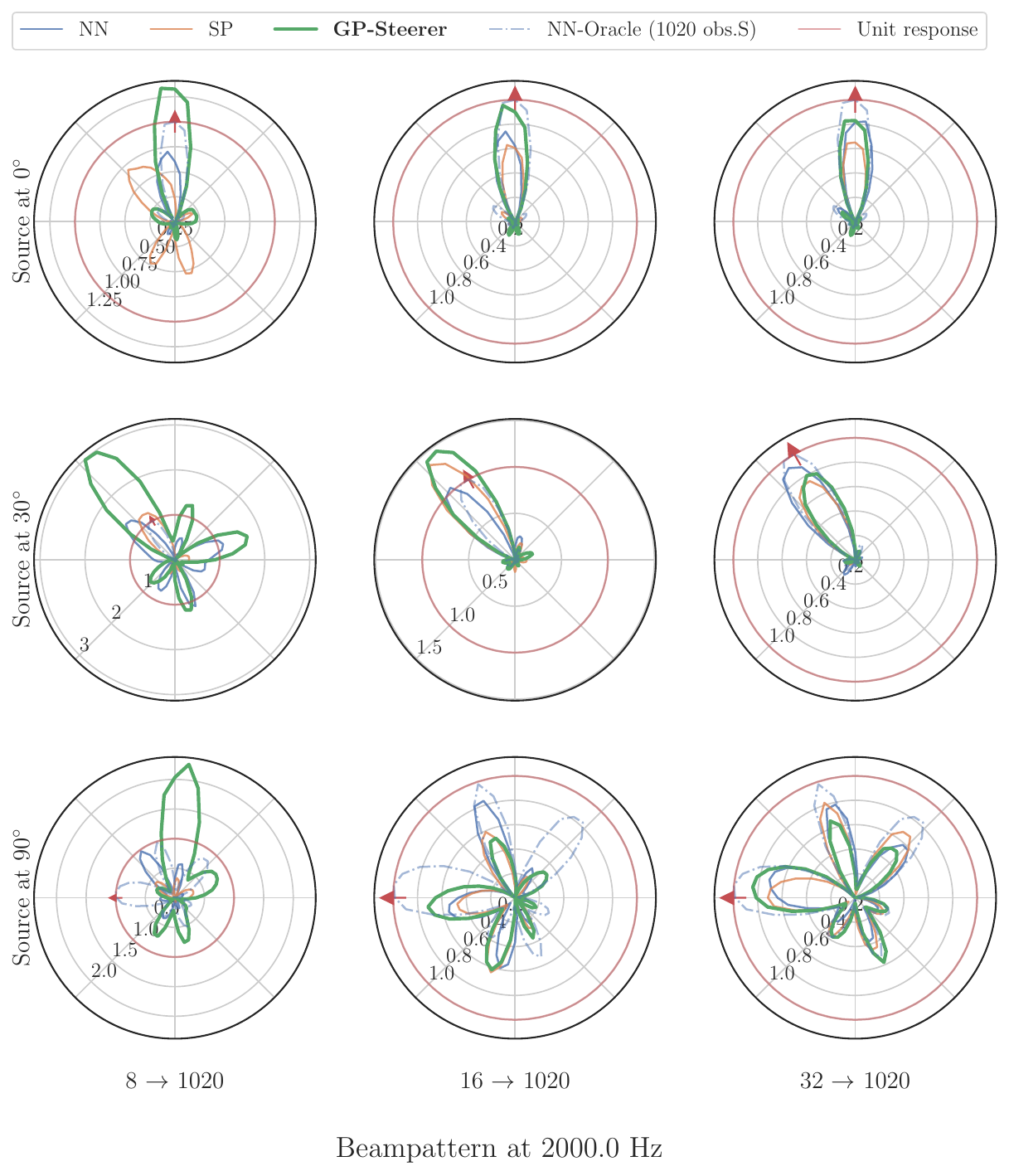}
    \caption{\small Beampatterns of a super-directive MVDR at the \SI{2}{\kilo\hertz}. Each panel corresponds to a different source direction and upsampling factor.}
    \label{fig:beampatterns}
    \vspace{-1em}
\end{figure}

\subsubsection{Experimental Evaluation}
This section analyzes 
 enhancement metrics across upsampling factors 
 in relation with the SV interpolation accuracy,
 and examines representative beampatterns at specific directions.

\cref{fig:metrics-correlation} reports
 the relationship between 
 interpolation fidelity, 
 measured in nMSE and CSIM, 
 and downstream speech enhancement performance across multiple metrics, 
 SDR, SAR, ISR, fwSegSNR, PESQ, MBSTOI. 
Each panel plots 
 the median enhancement improvement gains
 against the median interpolation performance.
For readability, 
 \SH, \GPChmat, and \NF are excluded
 due to underperformance in both tasks; 
 complete results are provided 
 in the supplementary materials.

In general,
 performances improve with increased spatial sampling. 
Moreover, for most metrics, 
 enhancement gains correlate strongly
 with interpolation quality, 
 underscoring the critical role 
 of steering vector accuracy 
 in downstream beamforming. 
Classic methods 
 like \NN and \SP show a 
 clear, monotonic improvement trajectory,
 approaching oracle-level performance 
 with only 128 observations. 
Notably, 
 a saturation effect is observed for \SP, 
 which reaches near-oracle performance with only 128 measurements. 
 \NN follows a similar, albeit slightly lower, trend. 
 
Concerning NF-based models, 
 \PCNN offers competitive results 
 with severe distortion-to-artifacts trade-off,
 whereas, \NFGW, \PINN exhibit
 greater variability and lack a consistent trend,
 likely due to sensitivity to initialization
 and hyperparameter tuning.
In particular, \PINN shows weak correlation, 
 pointing to difficulties in 
 balancing a multi-objective loss.

The proposed method \Proposed
 shows strong and consistent alignment
 between interpolation and enhancement quality for most metrics. 
However, 
its SDR and ISR trends are less stable, 
suggesting room for improvement 
in the underlying interpolation loss 
or its sensitivity to spectral balance
in the covariance model. 
In particular, small but incoherent high-frequency phase mismatches in the estimated SVs may propagate through the covariance estimate and translate into signal-level artifacts after beamforming, even when nMSE/CSIM improve.
Despite this, 
 the results confirm that
 combining GP regression 
 with physically structured kernels can significantly enhance 
 the generalization ability 
 of neural field approaches.
Moreover, 
 oracle performances in some metrics
 are achieved with very few measurements: 
 \Proposed outperforms \NNOracle in SDR, ISR and PESQ with $\Nobs=8$, and SAR with $\Nobs=8$.
Alas, surpassing oracle fwSegSNR and MBSTOI
 would require $\Nobs>128$.
 
Finally, 
 the strong performance of local methods,
 like \SP and \NN,
 may be attributed to the bias 
 toward frontal DOAs in the dataset: 
 19\% of segments involve targets within \SI{15}{\degree}, and nearly 50\% fall within \SI{25}{\degree}.
In this region, local interpolators perform nearly perfectly~(cfr.~\cref{fig:interp-adist-freq}), providing them with a structural advantage in real-world scenarios.

Azimuthal beampatterns 
 at \SI{2}{\kilo\hertz} 
 at different target directions 
 (\SI{0}{\degree}, \SI{30}{\degree}, and \SI{90}{\degree}), 
 for upsampling factor $\Nobs = 8, 16,32$ are shown in \cref{fig:beampatterns}. 
Almost all methods can reliably steer
 the beam towards the frontal direction
 even with few measured vectors, 
 confirming the preservation 
 of the frontal measurement. 
In contrast, 
 steering at \SI{90}{\degree} proves more challenging due to the lack of nearby training points and potential spatial aliasing. 
As $\Nobs$ increases, 
 spatial selectivity improves, 
 though some methods still exhibit spurious side lobes—likely due to array asymmetry. 
The beampatterns of \PINN are not shown due to significant inconsistency and out-of-scale response, hinting at a mismatch between physics‐based regularization and the actual data.

\section{Conclusion}\label{sec:conclusion}
We introduced a novel model that 
 combines Gaussian Process regression (GPR)
 with Neural Fields (NF)
 to effectively upsample 
 sparse steering vector (SV) measurements. 
Unlike prior knowledge-driven methods 
 that rely on rigid constraints 
 or complex regularization, 
 our method leverages a 
 physically motivated composite kernel
 and Bayesian inference 
 to achieve both flexibility and consistency.

Experimental results demonstrate
 that the proposed method outperforms
 classical interpolation 
 and recent deep learning techniques based on NF, 
 particularly at high frequencies 
 and in extrapolation scenarios, 
 where existing methods often fail. 
Crucially, 
 it maintains high fidelity to observed SVs,
 a key requirement for downstream robustness.
In speech enhancement tasks, 
 our model achieves near-oracle performance
 using less than 10\% of the original measurements.

While these findings are promising, 
 several avenues remain for future exploration.
First, 
 integrating more accurate physical models, 
 such as those derived from HRTF scattering analyses,
 may further improve kernel design. 
Second, 
 the computational complexity of 
 GPR with large covariance matrices 
 calls for scalable approximations, 
 already explored in the literature. 
Third, 
 the fixed geometry used here involved manually 
 tuned in-ear microphone positions, 
 which could be refined by modeling them 
 as latent variables within the GPR framework.
Finally, 
 although this study was based 
 on close-to-real simulations for objective evaluation, 
 perceptual validation through listening tests 
 is essential, particularly in light of the promising results in the SPEAR challenge context.
 
\bibliographystyle{IEEEtran}
\bibliography{IEEEabrv, AANabrv, ref}

\vfill

\end{document}

\section*{Supplementary Materials}

\begin{figure*}[h]
    \centering
    \includegraphics[width=0.9\linewidth]{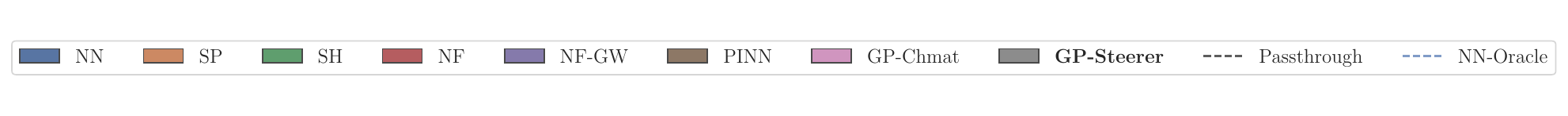}
    \includegraphics[trim={5 40 18 50},clip,width=0.49\linewidth]{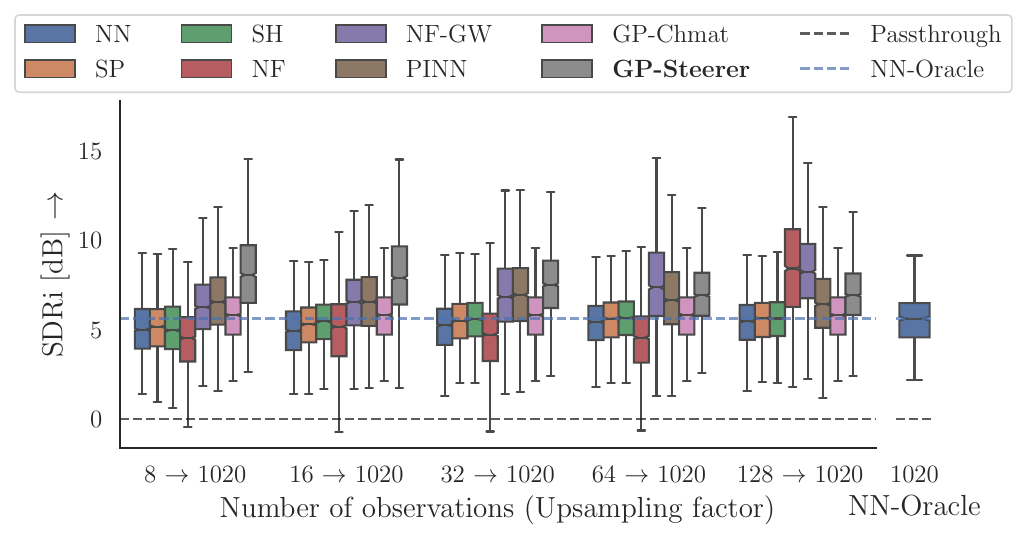}
    \includegraphics[trim={5 40 18 50},clip,width=0.49\linewidth]{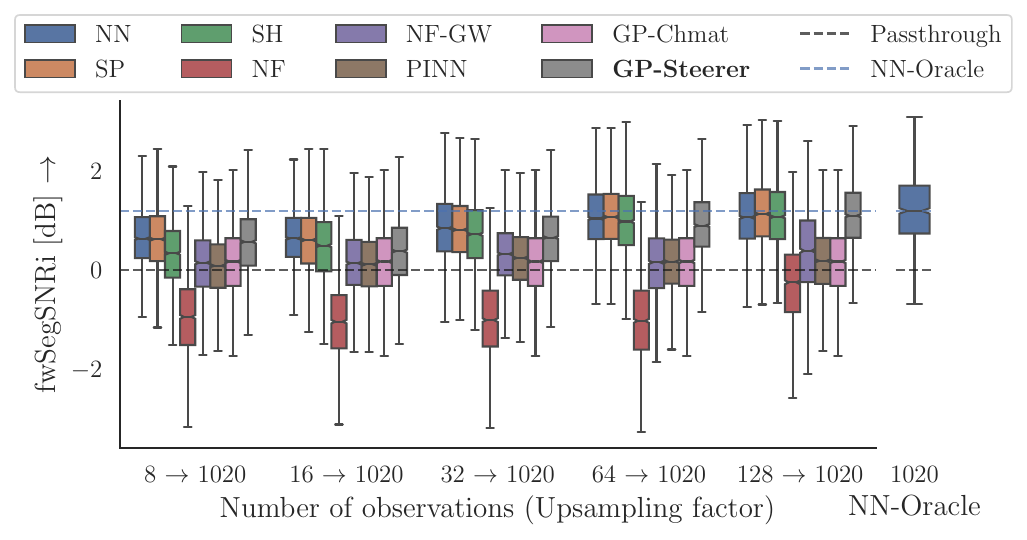}
    \\
    \includegraphics[trim={5 40 18 50},clip,width=0.49\linewidth]{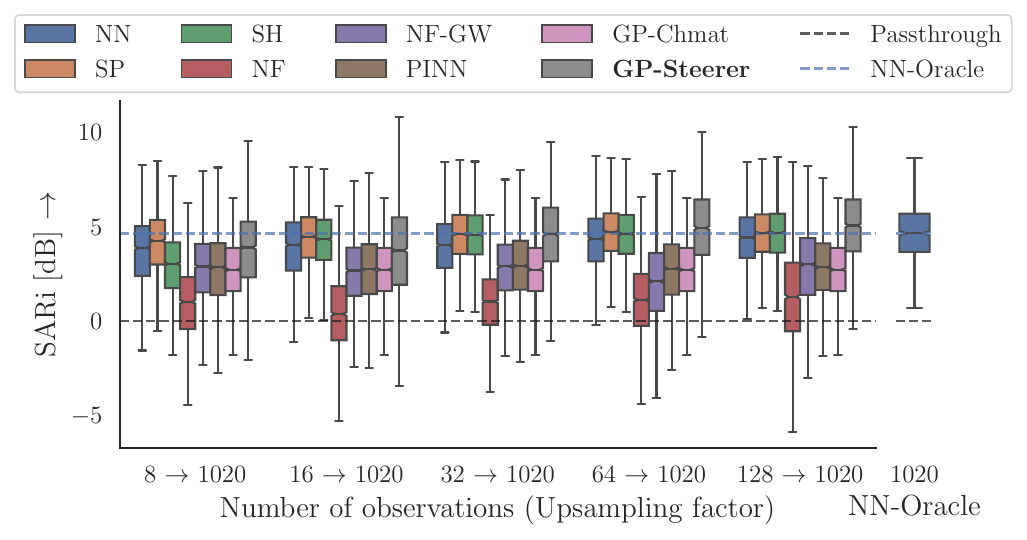}
    \includegraphics[trim={5 40 18 50},clip,width=0.49\linewidth]{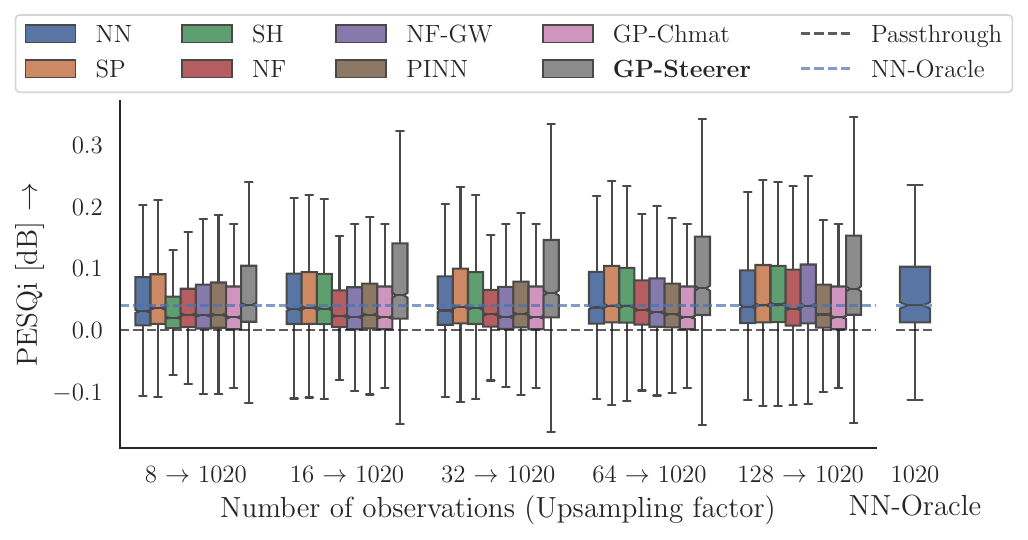}
    \\
    \includegraphics[trim={5 0 18 50},clip,width=0.49\linewidth]{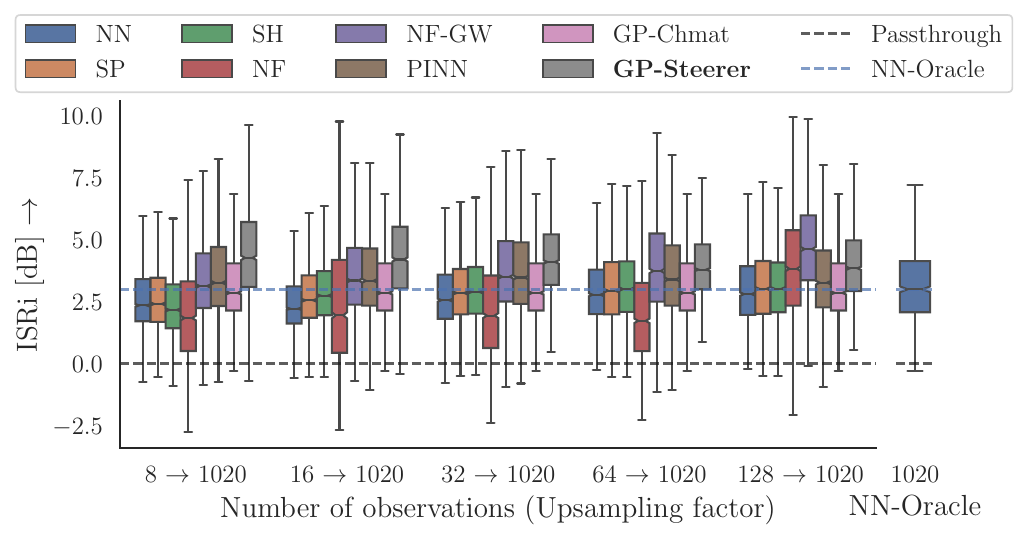}
    \includegraphics[trim={5 0 18 50},clip,width=0.49\linewidth]{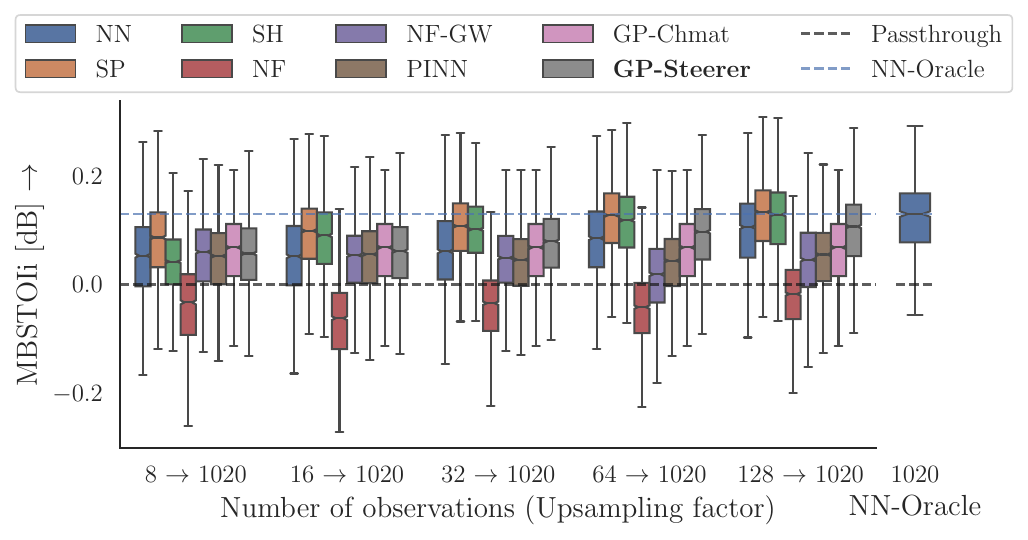}
    \caption{Enhancement results: improvements of energetic metrics (SDR, ISR, SAR, and fwSegSNR) in dB and perceptual metrics (PESQ, MBSTOI) relative to the passthrough per number of observed directions used to fit the corresponding upsampling model.}
    \label{fig:results-enhance}

\end{figure*}

\begin{table}[h]
    \centering
    \input{tables/model_count_params}
    \caption{\small Learnable-parameter counts per models and number of available observations $N_\text{obs}$. For \SH{} and \SP{} baselines, the number of parameters grows with the number of observations, whereas all Neural Fields-based models (\NF, \NFGW, \PINN, \PCNN) have constant parameter counts. $I, F, K$ are the number of sensors, frequencies, and SH coefficients, respectively.}
    \label{tab:placeholder}
\end{table}

\begin{figure}[h]
    \centering
    \includegraphics[width=0.8\linewidth]{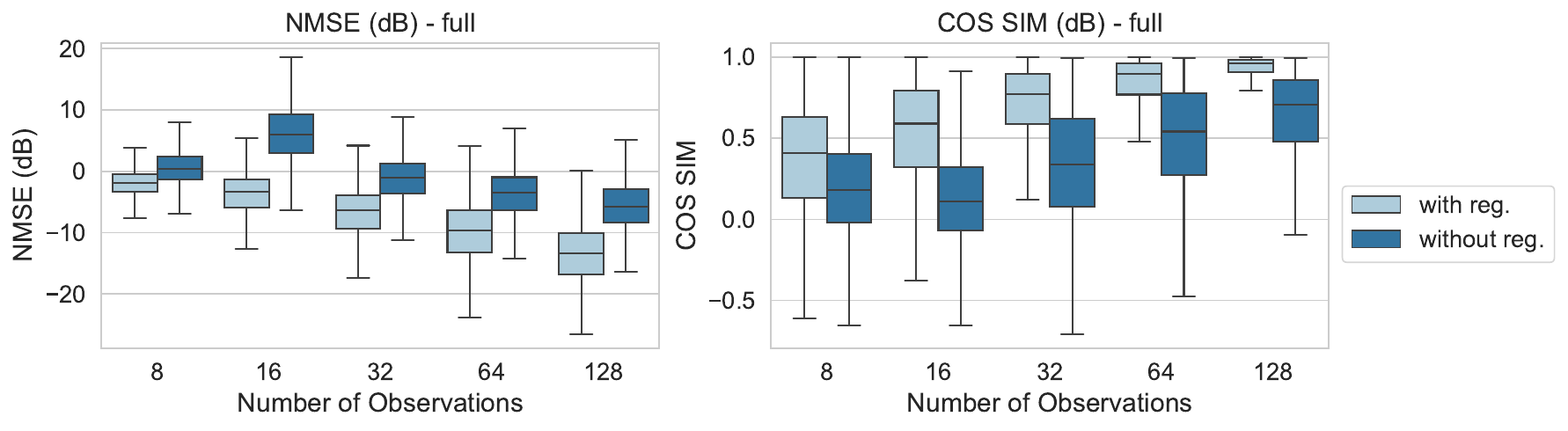}
    \caption{Ablation of the SH-spectrum regularizer in Eq.~(18) for the proposed model \Proposed. Boxplots show full-band SV reconstruction quality versus the number of observations $N_{\mathrm{obs}}$}
    \label{fig:placeholder}
\end{figure}

\begin{figure}[h]
    \centering
    \includegraphics[width=\linewidth]{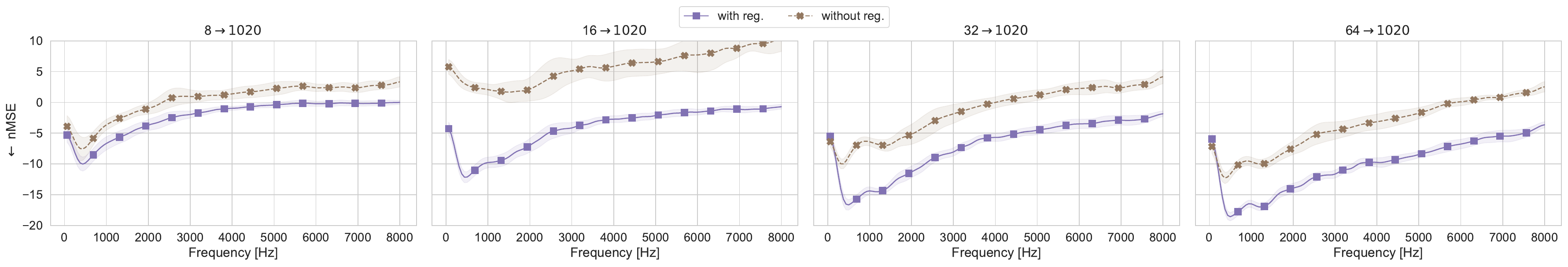}
    \caption{Ablation of the SH-spectrum regularizer in Eq.~(18) for the proposed model \Proposed for each frequency and three upsampling factor $N_\text{obs}$.}
    \label{fig:placeholder}
\end{figure}

\begin{figure}[h]
\centering
    \centering
    \hfill{}
    \subfloat[\label{fig:spear-speakers}]{{\includegraphics[width=0.25\linewidth]{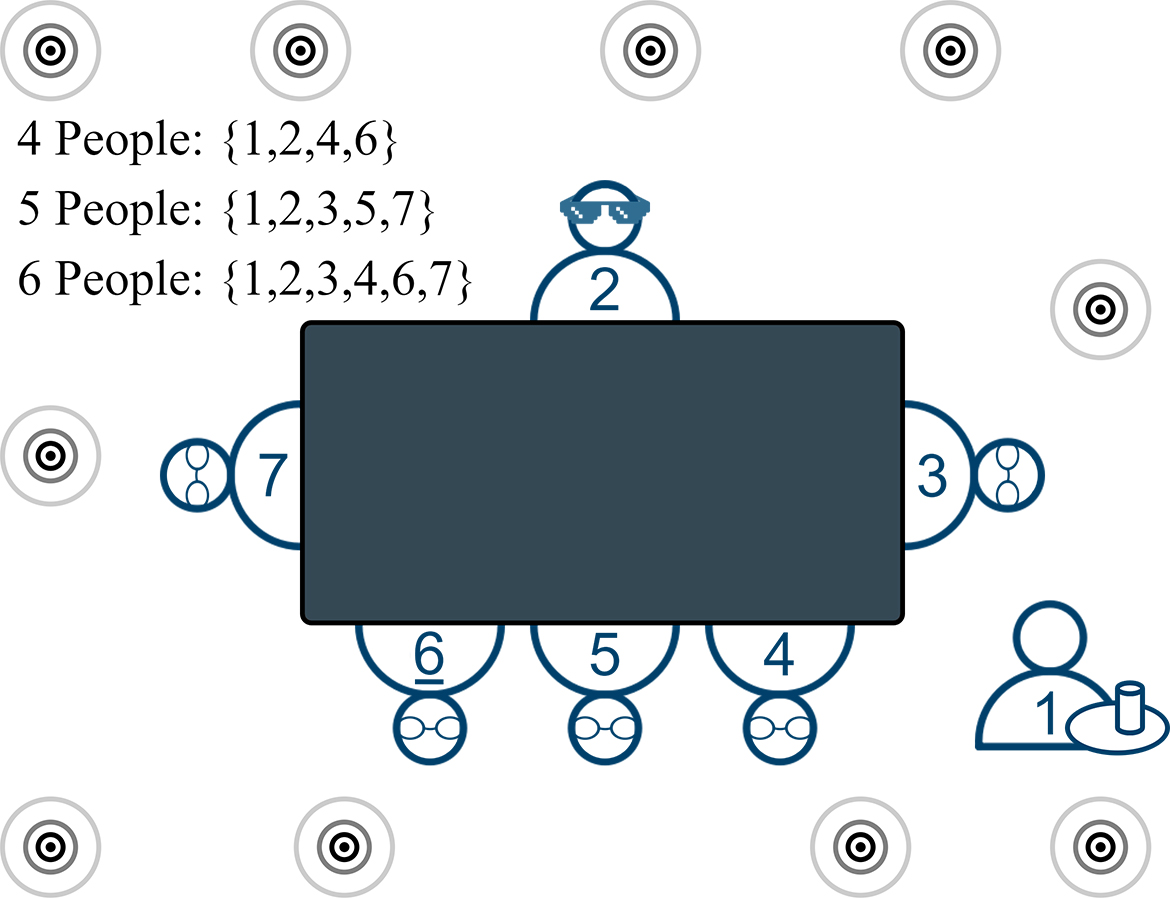} }}%
    \hfill
    \subfloat[\label{fig:spear-array}]
    {{\includegraphics[width=0.25\linewidth]{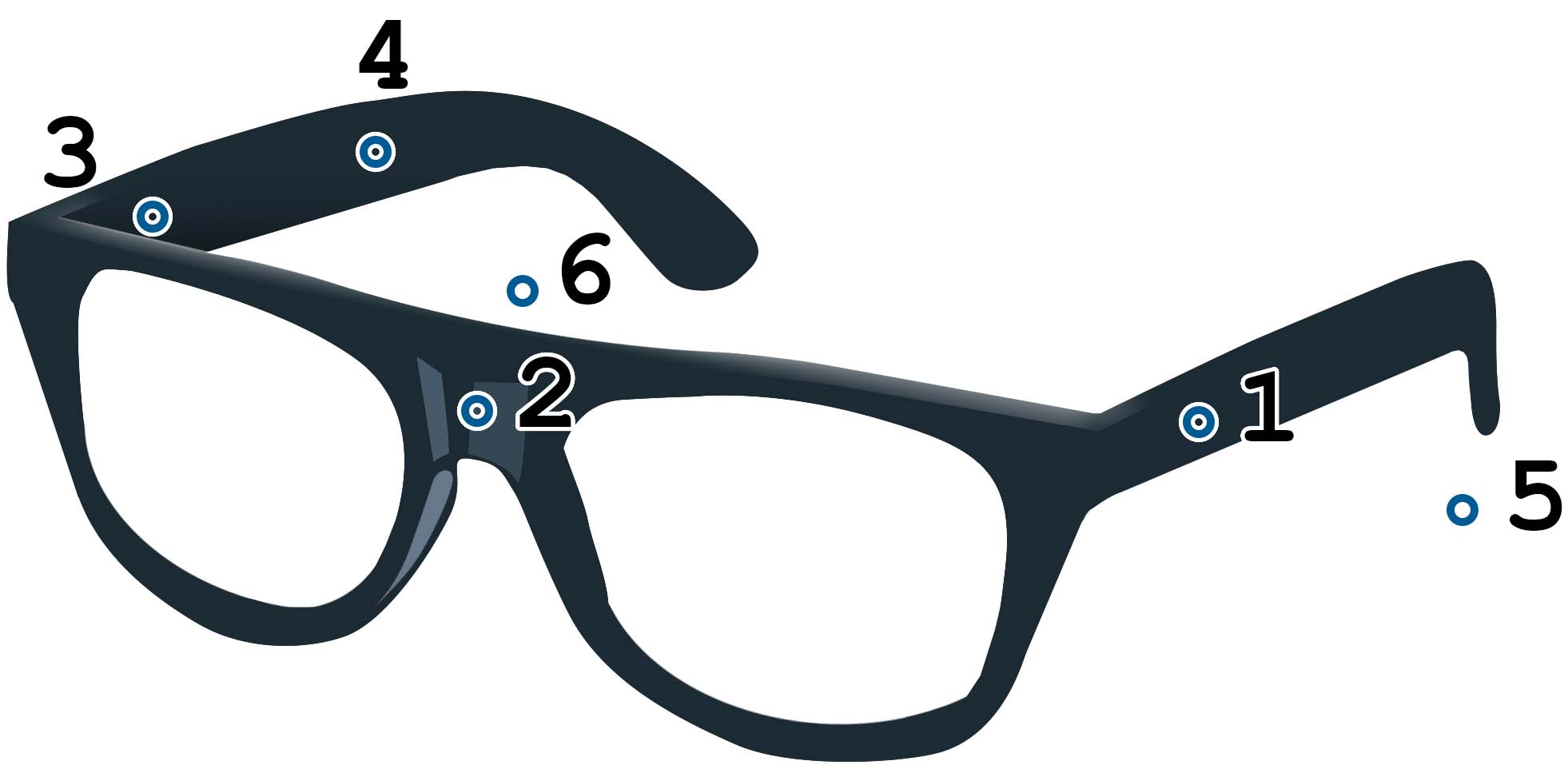} }}%
    \hfill{}
    
    \caption{The SPEAR Challenge (a) scene setup and (b) the microphone array worn by person number 2.  Image source: \url{https://github.com/facebookresearch/EasyComDataset}.}
    
\label{fig:spear-scene}
\end{figure}

\begin{table}[h]
    \centering
        \caption{Portion of target speaker DOAs in the SPEAR challenge data within an angular distance from frontal DOA $(\SI{0}{\degree},\SI{0}{\degree})$}
    \label{tab:spear_angles}
    \begin{tabular}{ccccccc}
    \toprule
    Angular Distance & \SI{5}{\degree} & \SI{10}{\degree} & \SI{15}{\degree} & \SI{20}{\degree} & \SI{25}{\degree} & \SI{30}{\degree}\\
    \midrule
    Portions target DOAs & 6\% & 19\% & 32\% & 43\% & 52\% & 61\% \\
    \bottomrule
    \end{tabular}
\end{table}

%% file: notation.tex
\newcommand{\anyPos}{\mathbf{q}}
\newcommand{\srcPos}{\mathbf{s}}
\newcommand{\micPos}{\mathbf{m}}
\newcommand{\arrPos}{\bar{{\mathbf{m}}}}

\newcommand{\sphBasis}{\psi}
\newcommand{\mbsphBasis}{\mbps}

\newcommand{\az}{\varphi}
\newcommand{\el}{\vartheta}
\newcommand{\Doas}{\Omega}

\newcommand{\Yml}{Y_l^m}

\newcommand{\param}{\boldsymbol{\theta}}
\newcommand{\mbZe}{\mathbf{0}}

\newcommand{\inv}[1]{{#1}^{-1}}

\newcommand{\NN}{\texttt{NN}\xspace}
\newcommand{\SH}{\texttt{SH}\xspace}
\newcommand{\SP}{\texttt{SP}\xspace}
\newcommand{\NF}{\texttt{NF}\xspace}
\newcommand{\NFGW}{\texttt{NF-GW}\xspace}
\newcommand{\PINN}{\texttt{PINN}\xspace}
\newcommand{\PCNN}{\texttt{PCNN}\xspace}
\newcommand{\GPChmat}{\texttt{GP-Chmat}\xspace}
\newcommand{\Proposed}{\texttt{GP-Steerer}\xspace}
\newcommand{\NNOracle}{\texttt{NN-Oracle}\xspace}

\newcommand{\Nobs}{N_\text{obs}}

%% file: notation_common.tex
\usepackage{amsmath,amsfonts,amssymb,bm}

\newcommand{\setR}{\mathbb{R}}
\newcommand{\setC}{\mathbb{C}}

\newcommand{\setS}{\mathbb{S}}

\newcommand{\Tr}{\mathsf{T}}
\newcommand{\Hr}{\mathsf{H}}

\newcommand{\mbC}{\mathbf{C}}

\newcommand{\mbI}{\mathbf{I}}

\newcommand{\mbK}{\mathbf{K}}

\newcommand{\mbR}{\mathbf{R}}

\newcommand{\mbW}{\mathbf{W}}

\newcommand{\mbZ}{\mathbf{Z}}

\newcommand{\mbb}{\mathbf{b}}
\newcommand{\mbc}{\mathbf{c}}
\newcommand{\mbd}{\mathbf{d}}

\newcommand{\mbg}{\mathbf{g}}
\newcommand{\mbh}{\mathbf{h}}

\newcommand{\mbk}{\mathbf{k}}

\newcommand{\mbm}{\mathbf{m}}

\newcommand{\mbw}{\mathbf{w}}
\newcommand{\mbx}{\mathbf{x}}
\newcommand{\mby}{\mathbf{y}}
\newcommand{\mbz}{\mathbf{z}}

\newcommand{\mbal}{\boldsymbol{\alpha}}

\newcommand{\mbga}{\boldsymbol{\gamma}}

\newcommand{\mbmu}{\boldsymbol{\mu}}

\newcommand{\mbps}{\boldsymbol{\psi}}

\newcommand{\mbSi}{\boldsymbol{\Sigma}}

\newcommand{\mbPs}{\boldsymbol{\Psi}}

%% file: tables/sota_table_small.tex
\begin{tabular}{
    p{0.03\linewidth}%
    p{0.05\linewidth}%
    p{0.3\linewidth}%
    p{0.22\linewidth}%
    p{0.25\linewidth}}%
\toprule
\multicolumn{3}{c}{}
    & \multicolumn{1}{l}{HRTF upsampling} 
    & \multicolumn{1}{l}{Sound Field Reconstr.} 
    \\
\midrule

\multicolumn{3}{c}{Recent reviews}
    & \multicolumn{1}{l}{\cite{porschmann2020comparison,bruschi2024review,cobos2022overview}}
    & \multicolumn{1}{l}{\cite{koyama2025physics,cobos2022overview}} 
    \\[0.1ex]
\hline\\[0.3ex]

\multirow{5}{*}{\rotatebox[origin=c]{90}{Data-driven}}      
        & \multicolumn{2}{l}{(Local) weighted interpolation}          
        & \multicolumn{1}{l}{\cite{porschmann2020comparison}} % HRTF
        & \multicolumn{1}{l}{$\sim$} % SFR
        \\
        
        & \multicolumn{2}{l}{(Global) Subspace  methods}
        & \multicolumn{1}{l}{\cite{xie2012recovery}}                % HRTF
        & \multicolumn{1}{l}{$\sim$}   % SFR 
        \\
                                                     
        & \multicolumn{2}{l}{Deep learning}           
        & \multicolumn{1}{l}{\cite{jiang2023modeling,gebru2021implicit,hogg2024hrtf}} % HRTF
        & \multicolumn{1}{l}{\cite{lluis2020sound,kristoffersen2021deep,pezzoli2022deep,miotello2024reconstruction}}     % SFR
        \\

        & \multicolumn{2}{r}{$\ldots$ with Gaussian Process}
        & \multicolumn{1}{l}{\cite{thuillier2024hrtf}} % HRTF
        & \multicolumn{1}{l}{\cite{liang2024sound}}     % SFR
        \\

        & \multicolumn{2}{r}{$\ldots$ with multimodalies}    
        & \multicolumn{1}{l}{\cite{chen2019autoencoding}}                     % HRTF
        & $\sim$
        \\
        & \multicolumn{2}{r}{$\ldots$ with Neural Fields}
        & \multicolumn{1}{l}{\cite{gebru2021implicit,zhang2023hrtf}} % HRTF
        & \multicolumn{1}{l}{$\sim$}     % SFR
        \\[3pt]
        
        & \multicolumn{2}{l}{Manifold Learning}
        & \multicolumn{1}{l}{\cite{duraiswami2005manifolds,deleforge2015acoustic,grijalva2017interpolation}} % HRTF
        & \multicolumn{1}{l}{$\sim$}    % SFR
        \\[2pt] \hline

\multirow{7}{*}{\rotatebox[origin=c]{90}{Knowledge-driven}} 
    & \multicolumn{2}{l}{Geometric-based}                 
        & \multicolumn{1}{l}{\cite{pulkki1997virtual,luo2013kernel,chen2023head,zotkin2009regularized}} % HRTF
        & $\sim$
        \\
                                  
        & \multicolumn{2}{l}{Parametric (DSP)-based}          
        & \multicolumn{1}{l}{\cite{watanabe2003interpolation,nowak2022spatial,lee2023neural,masuyama2024niirf}}
        & $\sim$
        \\[3pt]

        & \multirow{4}{*}{\rotatebox[origin=c]{90}{Physics-{\scriptsize based}}} 
        & Physics-constrained 
        & \multicolumn{1}{l}{\cite{evans1998analyzing,duraiswami2004interpolation,zotkin2009regularized,ahrens2012hrtf,porschmann2019directional,ben2019efficient,arend2021assessing}}
        & \multicolumn{1}{l}{\cite{bertin2015compressive,koyama2019sparse,damiano2021soundfield,caviedes2021gaussian,ribeiro2024sound,pezzoli2022sparsity,das2021room,antonello2017room}}%
        \\

        & 
        & \multicolumn{1}{r}{$\ldots$ with Deep Learning}
        & \multicolumn{1}{l}{\cite{ito2022head}}      
        & \multicolumn{1}{l}{\cite{karakonstantis2023generative,ribeiro2024sound,bi2024point}}
        \\

        & 
        & \multicolumn{1}{r}{$\ldots$ with Gaussian Process}
        & \multicolumn{1}{l}{\cite{luo2013kernel,romigh2015bayesian}}
        & \multicolumn{1}{l}{\cite{caviedes2021gaussian,caviedes2023spatio,feng2024room}}
        \\
        
        &                                
        & Physics-informed    
        & \multicolumn{1}{l}{\cite{ma2024sound}}
        & \multicolumn{1}{l}{\cite{pezzoli2023implicit,ma2024sound,karakonstantis2024room}}  
        \\
        
        & & 
        & \multicolumn{1}{l}{}               
        & \multicolumn{1}{l}{}  
        \\
    
        \bottomrule
\end{tabular}

%% file: tables/model_count_params.tex
\begin{tabular}{@{}cc|cccccc@{}}
\toprule
    & & & \multicolumn{5}{c}{Upsampling factor $N_\text{obs}\to$ 1020} \\
Model
    % & Stored input? 
    & Number of learnable parameters                    
    & 8
    & 16                           
    & 32                           
    & 64                           
    & 128                           
    \\ \midrule
input size 
    % &
    & $I \times F \times N_\text{obs}$
    & 10320 
    & 20640 
    & 41280 
    & 82560
    & 165120 
    \\ \midrule
\NN         
    % & Yes           
    &                               
    & N/A                          
    & N/A                          
    & N/A                          
    & N/A                          
    & N/A                          
    \\
\SH         
    % & No            
    & $I \times K \times F$
    & 3096                         
    & 12384                        
    & 19350                        
    & 49536                        
    & 93654                         
    \\
\SP         
    % & No            
    & $I \times F \times N_\text{obs}$
    & 6192                         
    & 12384                        
    & 24768                        
    & 49536                        
    & 99072                         
    \\
\NF         
    % & No            
    & PE + MLP                      
    & 34306                        
    & \%                           
    & \%                           
    & \%                           
    & \%                            
    \\
\NFGW      
    % & No            
    & PE + MLP                      
    & 34306                        
    & \%                           
    & \%                           
    & \%                           
    & \%                            
    \\
\PINN       
    % & No            
    & PE + MLP                      
    & 34306                        
    & \%                           
    & \%                           
    & \%                           
    & \%                            
    \\
\PCNN
    % & No            
    & PE + MLP                      
    & 59848                        
    & \%                           
    & \%                           
    & \%                           
    & \%                            
    \\
\GPChmat   
    % & Yes           
    & Kernel hyperparams            
    & 5                            
    & \%                           
    & \%                           
    & \%                           
    & \%                            
    \\
\Proposed{}
    % & Yes           
    & Kernel hyperparams + PE + MLP 
    & 88312                        
    & \%                           
    & \%                           
    & \%                            
    & \%                            
    \\ \bottomrule
\end{tabular}